\documentclass[pre,floatfix,longbibliography,superscriptaddress,twocolumn,final,notitlepage]{revtex4-2}

\RequirePackage{amsmath}
\RequirePackage{hyperref} 
\RequirePackage{amssymb}
\RequirePackage{graphicx}
\RequirePackage{bbm}
\RequirePackage{xcolor}
\usepackage{graphicx}
\usepackage{subcaption}
\usepackage{amsmath}
\usepackage{natbib}
\usepackage{upgreek}
\usepackage{caption}
\usepackage[nameinlink,capitalize]{cleveref}
\usepackage{amsthm}
\usepackage{placeins}
\usepackage{pifont}
\usepackage{booktabs}
\usepackage{nicefrac}

\def\andname{\hspace*{-0.5em}}
\setcitestyle{round}
\usepackage{makecell}

\newcommand{\beginsupplement}{
        \setcounter{table}{0}
        \setcounter{equation}{0}
        \renewcommand{\thesection}{\Alph{section}}
	\renewcommand{\thesubsection}{\thesection.\arabic{subsection}}
	\renewcommand{\theequation}{S\arabic{equation}}
 }

\newcommand{\Exp}{\mathbb{E}}
\newcommand{\be}{\begin{equation}}
\newcommand{\ee}{\end{equation}}
\newcommand{\bea}{\begin{eqnarray}}
\newcommand{\eea}{\end{eqnarray}}

\newcommand{\f}[2]{\frac{#1}{#2}}

\newcommand{\ccup}[1]{\left\{#1\right\}}

\newcommand{\m}{\theta}
\newcommand{\vm}{\boldsymbol{\theta}}

\newcommand{\bs}{\boldsymbol}
\newcommand{\ms}{\mathsmaller}
\newcommand{\e}{\edgetype}
\newcommand{\Ad}{A^{\mathsmaller{(d)}}}
\newcommand{\Aall}{\mathcal{A}^{\mathsmaller{(:)}}}
\newcommand{\Lambdad}{\mathrm{\Lambda}^{\mathsmaller{(d)}}}
\newcommand{\Lambdaall}{\Uplambda^{\mathsmaller{(:)}}}
\newcommand{\Omegad}{\Omega^{\ms{(d)}}}

\newcommand{\Capmu}{{\mathlarger{\mu}}}
\newcommand{\Mud}{\Capmu^{\ms{(d)}}}
\newcommand{\Muall}{\mathlarger{\upmu}^{\ms{(:)}}}

\newcommand{\edgetype}{\mathbf{i}}
\newcommand{\idxrange}[2]{{#1}_{1}\dots{#1}_{#2}}
\newcommand{\iidx}{\idxrange{i}{d}}
\newcommand{\cidx}{\idxrange{c}{d}}

\usepackage{algorithm, algpseudocode}
\usepackage{amsmath,fixmath}
\usepackage{fixmath}
\usepackage{setspace}
\usepackage{float}
\usepackage{relsize}
\usepackage{stmaryrd}

\theoremstyle{plain}
\newtheorem{theorem}{Theorem}[section]

\newtheorem{lemma}[theorem]{Lemma}
\newtheorem{corollary}[theorem]{Corollary}
\theoremstyle{definition}
\newtheorem{definition}[theorem]{Definition}

\theoremstyle{remark}

\usepackage{booktabs}
\usepackage{adjustbox}
\renewcommand{\thetable}{\arabic{table}}
 
\definecolor{shadecolor}{gray}{0.9}
\definecolor{maroon}{RGB}{128, 0, 0}
\definecolor{bronze}{RGB}{205, 127, 50}
\usepackage{tabu}

\usepackage[normalem]{ulem}
\usepackage{soul}

\usepackage{soul}

\newcommand{\algo}{\mbox{{\textsc{Omni-Hype-SMT}}}}

\usepackage{pifont}

\newcommand{\sprod}{\mathop{\mathsmaller{\prod}}}

\begin{document}

\title[Broad Spectrum Structure Discovery in Large-Scale Higher-Order Networks]{Broad Spectrum Structure Discovery in Large-Scale Higher-Order Networks}

\twocolumngrid

\author{John Hood}
	\email{johnhood@uchicago.edu}
	\affiliation{Department of Statistics, University of Chicago}

\author{Caterina De Bacco}
    \email{c.debacco@tudelft.nl}
	\affiliation{Department of Quantum and Computer Engineering, Delft University of Technology}

\author{Aaron Schein}
\email{schein@uchicago.edu}
	\affiliation{Department of Statistics, University of Chicago}

\begin{abstract} 

Complex systems are often driven by higher-order interactions among multiple units, naturally represented as hypergraphs. 
Understanding dependency structures within these hypergraphs is crucial for understanding and predicting the behavior of complex systems but is made challenging by their combinatorial complexity and computational demands. In this paper, we introduce a class of probabilistic models that efficiently represents and discovers a broad spectrum of mesoscale structure in large-scale hypergraphs. The key insight enabling this approach is to treat classes of similar units as themselves nodes in a latent hypergraph. By modeling observed node interactions through latent interactions among classes using low-rank representations, our approach tractably captures rich structural patterns while ensuring model identifiability. This allows for direct interpretation of distinct node- and class-level structures. Empirically, our model improves link prediction over state-of-the-art methods and discovers interpretable structures in diverse real-world systems, including pharmacological and social networks, advancing our ability to incorporate large-scale higher-order data into the scientific process.~\looseness=-1
\end{abstract}

\maketitle

\section{Introduction} 
Complex systems---social, biological, or technological, among other types---are often driven by \textit{higher-order interactions} among potentially many nodes~\cite{battiston2020networks}. Such systems can be modeled as \textit{hypergraphs}, which extend the traditional notion of graphs or networks from dyadic or pairwise interactions to those of higher order. 

Like traditional networks, real-world hypergraphs exhibit \textit{mesoscale structure} \cite{porter2009communities,fortunato2010community}---i.e., patterns of interaction among groups of nodes. Broadly, modeling such structure reduces to clustering nodes into groups and characterizing how those groups interact, if at all. In doing so, one reduces the conceptual complexity of the system and the dimensionality of the data, potentially revealing real functional components and underlying mechanisms that drive observed interactions~\cite{antelmi2023survey}.

Mesoscale structure comes in many different forms, not all of which are tractably modeled. Perhaps the most commonly modeled structure is \textit{assortative} structure wherein nodes form ``communities'' and interact mostly with other similar nodes within the same community. Recent work develops methods for efficient detection of these communities in hypergraph data~\cite{chodrow2021generative, contisciani2022inference,ruggeri2023community}. However, many complex systems also exhibit some degree of \textit{disassortative} structure, wherein similar nodes form ``classes'', but may not interact within these classes. Within these classes, nodes have similar characteristics, but may (exclusively) interact with nodes in other classes. Predator-prey networks in ecology are one such example, wherein species take one of two classes (predator or prey) and interact mostly with species of the other. ~\looseness=-1

Even in traditional network settings, any degree of disassortativity (interaction between nodes of different classes) is generally difficult to model. This challenge stems from the baseline complexity required to model both how nodes form groups and how these combinations of groups interact. For hypergraphs, this complexity compounds: higher-order interactions among nodes introduce higher-order interactions within and between groups, resulting in a combinatoric explosion in the number of model parameters that makes estimation impossible. Recent work contends with this by developing models that surface highly restricted forms of disassortativity in hypergraphs, such as structure that can be modeled by a Bethe approximation~\cite{ruggeri2023community}, or \textit{core-periphery} structure wherein a dense
core of tightly connected nodes can be distinguished from a sparse loosely connected periphery of nodes~\cite{papachristou2022core,tudisco2023core}.~\looseness=-1  

\begin{figure*}[t!]
    \centering
\includegraphics[width=\linewidth]{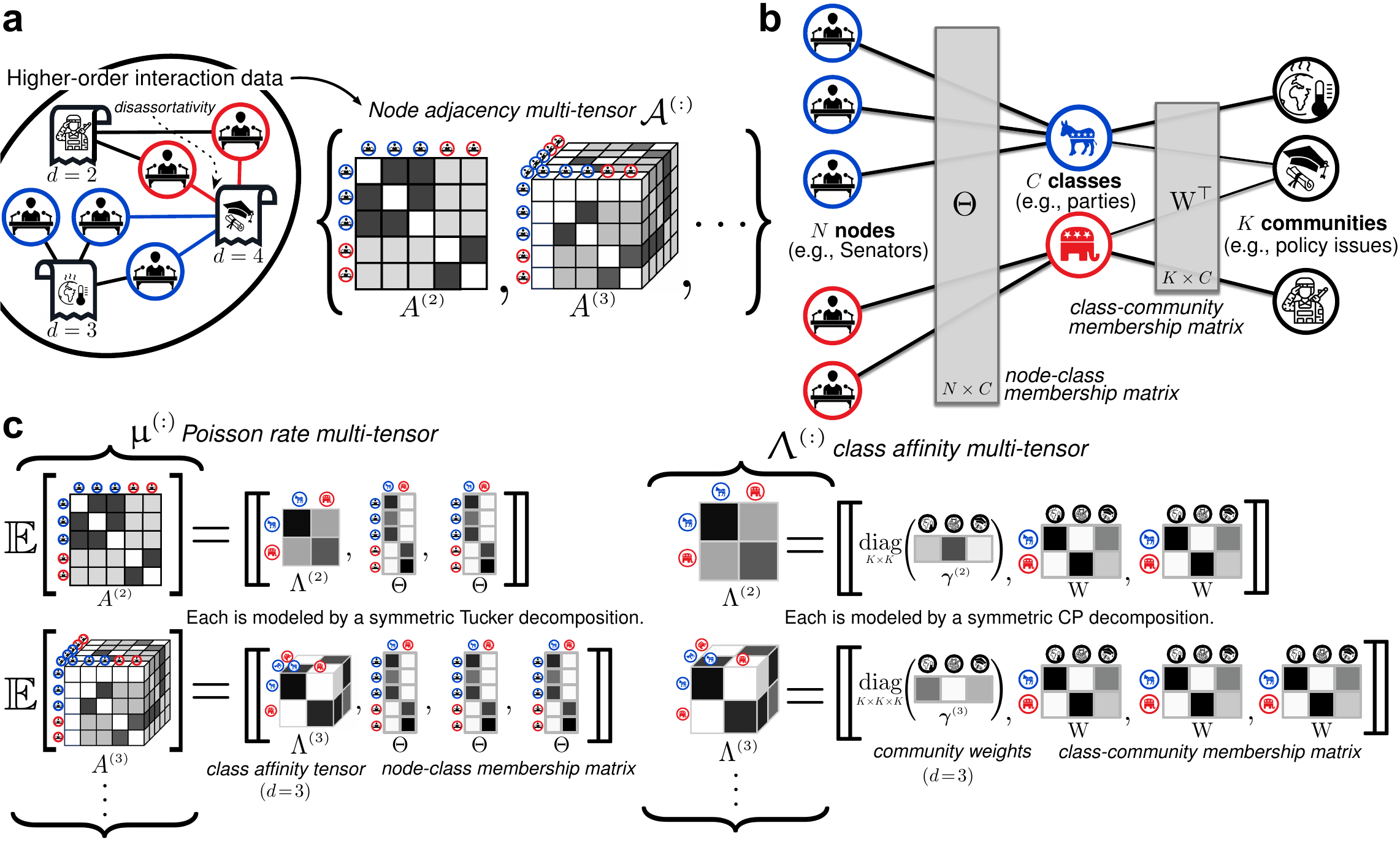}
\caption{\textbf{\algo~models a range of mesoscale structures by 
 assigning nodes to classes and classes to communities.} A schematic illustration detailing how \algo~models a diverse range of mesoscale structure. a) Hypergraph data encodes higher-order interactions among entities, such as bill co-sponsorship (left). Each hyperedge captures a multi-way relationship among nodes. These interactions are represented as a multi-order adjacency tensor $\Aall$ (right), where $\Ad$ denotes the order-$d$ adjacency tensor capturing $d$-way interactions. Such a formulation enables the modeling of complex, non-pairwise relational structures. b) Two parameters $\mathrm{\Theta}$ and $\mathrm{W}$ govern interactions between nodes (e.g., Senators). The node-class membership matrix $\mathrm{\Theta}$ soft clusters similar nodes to classes (e.g., political parties), while the class-community membership matrix $\mathrm{W}$ models interactions between classes through communities (e.g., policy issues). Together, the matrix $\mathrm{\Theta} \mathrm{W}$ captures assortative and disassortative activity between nodes through interactions within and between classes. c) The multi-tensor $\Muall$ (left) models the observed higher-order adjacency tensors $\Aall$ via symmetric low-rank tensor decomposition using a class affinity tensor $\Lambdad$ and a node-class membership matrix $\mathrm{\Theta}$, shared over orders $d$. The class affinity multi-tensor $\Lambdaall$ (right) is further decomposed into community-order rates $\gamma^{\ms{(d)}}_k$ and the symmetric outer products of the columns of class-community membership matrix $\mathrm{W}$, enabling interpretable modeling of multi-way class interactions across different orders.}
    \label{fig:toy}
\end{figure*}

While prior efforts have analyzed a wide range of hypergraph settings theoretically \cite{chodrow2021generative,sales2023hyperedge,ruggeri2024message,brusa2024model,veldt2022hypergraph,pister2024stochastic}, practitioners are limited to only a few options for modeling mesoscale structure in hypergraphs tractably, each of which has its own drawbacks. One approach restricts analysis to assortative (or similarly constrained) structure, risking mischaracterization of a non-assortative system. Another limits the data to a moderate number of small-order interactions (e.g., three- or four-way) to enable the application of one of the more flexible existing approaches which scale poorly to large, high-order hypergraphs. A third strategy reduces higher-order data to pairwise interactions, potentially discarding crucial structural information through data preprocessing steps~\cite{yoon2020much,contisciani2022inference,purkait2016clustering}. Collectively, these limitations significantly hamper researchers' ability to leverage large-scale data to draw reliable conclusions regarding the complex systems they study.~\looseness=-1

Motivated by these challenges, this paper introduces a family of probabilistic generative models to tractably capture a wide range of mesoscale structure underlying large-scale hypergraphs. This family encompasses several existing models and spans an \textit{omniassortative} spectrum of structural patterns, ranging from strictly assortative to highly disassortative. Where along this spectrum a given model falls is dictated by its parameter values.~\looseness=-1

The key idea driving our approach is to jointly cluster nodes into \textit{classes} and classes into \textit{communities}. Each clustering is soft, permitting nodes to belong to multiple classes, and classes to multiple communities, as shown in~\Cref{fig:toy}b. The proposed model, which we refer to as \algo, avoids the combinatoric explosion associated with modeling all combinations of inter-class interaction by dictating that classes interact exclusively within communities. By clustering classes into communities, \algo~allows for disassortative interactions between nodes. Since classes are made up of similar nodes, interactions between different classes yield disassortative interactions between nodes of those classes. It is through this framework that the proposed model captures rich and interpretable latent structure governing higher-order interactions among nodes. ~\looseness=-1 

We leverage a principled framework of statistical inference to rigorously identify mesoscale structure in hypergraph data. Our proposed model is defined such that its parameters are provably identifiable from the data, enhancing the reliability and robustness of their interpretation. Beyond providing a rigorous proof of model identifiability, we derive efficient parameter updates by exploiting the probabilistic nature of the proposed model and conditional distributions within.  It is this efficiency that enables large-scale, data-driven discovery of mesoscale structure in practice. We demonstrate how this efficiency enables the development of an extremely simple yet scalable algorithm for generating synthetic hypergraphs with tunable mesoscale structure—a capability largely missing from the current literature~\cite{chodrow2021generative, ruggeri2024framework}, due to the computational challenges posed by the high-dimensional nature of hypergraphs. Altogether, the analytic and computational tractability of our approach, along with its theoretical guarantees, makes it a principled and effective way to disentangle different types of mesoscale structure underlying large-scale complex systems.~\looseness=-1

We demonstrate the expressiveness and scalability of \algo~through extensive experiments on two biological datasets, three human-contact networks, and three political datasets, each of which has a natural hypergraph representation. We find a diverse range of mesoscale structure extending beyond strict assortativity. We prove that a state-of-the-art assortative model~\cite{contisciani2022inference} is a specific, restricted instance of the proposed model class. We focus our comparisons here to demonstrate the proposed model's advanced capabilities. Our flexible modeling approach leads to better performance on downstream tasks, yielding enhanced higher-order link prediction and more interpretable node clustering than existing approaches.~\looseness=-1

Our results reflect an intuitive insight: as the order of interaction increases, appropriately modeling disassortative structure becomes more important. In a case study of higher-order drug combinations among emergency room (ER) patients, we demonstrate how learning diverse types of mesoscale structure provides more nuanced insights into drug classes and their interaction patterns. For example, pairwise interactions between drugs tend to be between ``recreational drugs''---a class inferred by the model and labeled post hoc via generative AI---such as alcohol and marijuana. On the other hand, among patients with many drugs in their system, if one is a ``cardiovascular medication'' such as lisinopril or metoprolol, the others are often ``psychotropic medications'' (quetiapine, clonazepam) or ``opioid analgesics'' (oxycodone, hydrocodone, fentanyl). When mixed, these classes of drugs that are often prescribed for medical benefit increase in potency. These insights may inform clinical risk assessment and the design of safer polypharmacy protocols by revealing latent, higher-order interaction patterns that are obscured in pairwise analyses. ~\looseness=-1

Together, these findings highlight the importance of modeling a full spectrum of mesoscale structure in uncovering clinically and scientifically meaningful patterns in complex, higher-order data. ~\looseness=-1

\section{Results}

\subsection{Motivating departure from strict assortativity: two case studies}
We start by showing how existing probabilistic approaches, which are based on assumptions of strict assortativity, provide sub-optimal representations of the data when fit to hypergraphs exhibiting realistic degrees of disassortativity. We analyze two datasets, each exhibiting a different type of mesoscale structure departing from strict assortativity.~\looseness=-1

First, we draw upon a dataset of human-contact interactions in a hospital, where nodes are either staff (i.e., doctors, nurses, or administrative assistants) or patients~\cite{vanhems2013estimating}. In this dataset, hyperedges describe proximity interactions as measured by wearable Bluetooth devices. We create a semi-synthetic version of the data by removing hyperedges that do not contain at least one patient. Thus, all interactions are either exclusively between patients or those consisting of at least one patient and one staff. We then fit two models: 1) the state-of-the-art strictly assortative model~\cite{contisciani2022inference} mentioned above, and 2) our omniassortative model.~\looseness=-1

\begin{figure*}[t!]
    \centering
    \includegraphics[width=\linewidth]{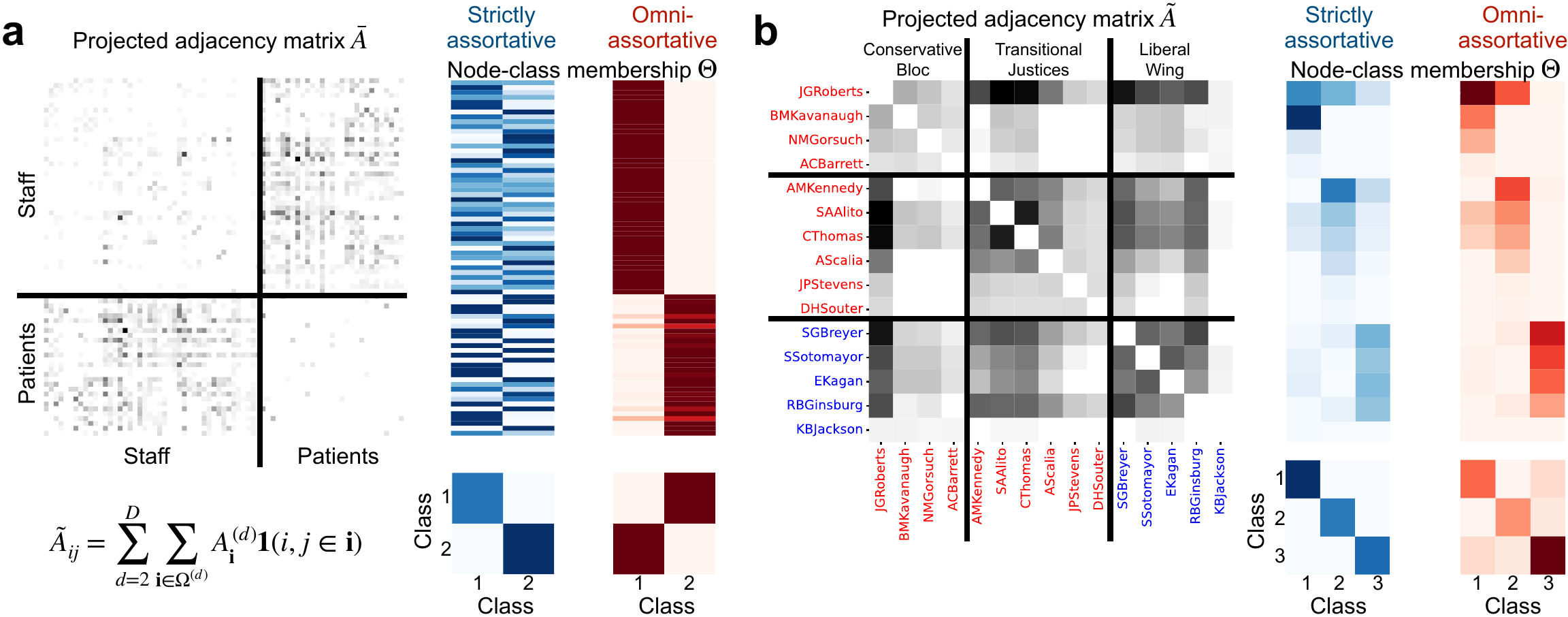}
    \caption{\textbf{\algo~recovers core-periphery structure among US Supreme Court justices and disassortative structure in hospital proximity data.} In each setting, the projected adjacency matrix of the hypergraph data is visualized next to the inferred node-class membership and class affinity matrices. The omniassortative model's class affinity matrix has non-zero intensity on its off-diagonal, representing possible interactions between classes, while the strictly assortative matrices are zero on the off-diagonal. a) In the hospital setting, the node-class membership matrix cleanly separates patients from staff for the omniassortative model but not the strictly assortative one; we normalize each node's membership vector to emphasize this difference. For the omniassortative model, the class affinity matrix is strongly off-diagonal, suggesting that interactions among patients and staff are highly disassortative. b) In the Supreme Court Justice setting, we color-code the justices according to the party of the nominating President, and see that both models cleanly separate three blocks of justices, known loosely as the liberals, conservatives, and transitional justices. The omniassortative model's node-class membership matrix is sharper, with less mixed-membership, and it explains the crossover in voting patterns across blocks by inferring core-periphery structure, where class affinity matrix is strongly diagonal but has a non-negligible off-diagonal.~\looseness=-1}
    \label{fig:hospital}
\end{figure*}

At a high-level, both models can be understood as soft-clustering $N$ nodes into $C$ latent classes, represented by an $N \times C$ node-class membership matrix. 
We visualize the matrices learned by each of the two models in the bottom panel of ~\Cref{fig:hospital}a. 
The strictly assortative model (\textit{left column}) requires that nodes of different classes do not interact and fails to recover the underlying staff-versus-patient block structure. On the other hand, the omniassortative model (\textit{right column}) cleanly partitions the nodes into patients and staff and appropriately models interactions between the groups. 
In addition to providing a more interpretable group-level description of the data, the omniassortative model predicts heldout hyperedges better than the strictly assortative one (AUC of 0.91 versus 0.85).~\looseness=-1

In our second example, we use a dataset of United States Supreme Court cases ranging over the years 2005--2024. 
The data forms a hypergraph wherein nodes are Supreme Court justices and each hyperedge corresponds the set of justices who concurred with the majority opinion of a given case. 
Here, both models recover similar block structure, shown in the top panel of \Cref{fig:hospital}b. 
However, the classes inferred by the assortative model are less distinguishable, as measured by the entropy $\mathbb{H}(\mathrm{\Theta})$ (defined in \Cref{sec:application}) of the node-class membership matrix. 
The omniassortative model, on the other hand, cleanly distinguishes between classes corresponding largely to Democrat- versus Republican-appointed justices. 
The strictly assortative model assigns the Republican-appointed Justices Roberts, Kennedy, Alito, Thomas, and Scalia partial membership in a class dominated by Democratic appointees, while the omniassortative model does not. 
The omniassortative model permits interactions between classes---thus, it does not require heavily mixed class membership to explain interactions between Republican-appointed justices who sometimes concur with Democrat-appointed ones. 
The two bottom-most panels in \Cref{fig:hospital}b depict affinities between classes, where we note that the strictly assortative affinity matrix (blue) is constrained to be diagonal.~\looseness=-1

These examples represent two distinctly different departures from strict assortativity. 
The first shows highly disassortative structure, where patients and staff primarily interact across classes, instead of within them. 
The second example shows core-periphery structure, where justices primarily agree with other justices of similar ideology (i.e., conservatives agree with conservatives) but also agree with justices of opposing ideologies (such as in unanimous rulings) albeit less often. 
These simple examples motivate our modeling approach, described next.~\looseness=-1

\subsection{\algo: an \underline{\textit{omni}}assortative model of \underline{\textit{hype}}rgraphs from \underline{\textit{s}}ymmetric \underline{\textit{m}}ulti-\underline{\textit{t}}ensors}

Here, we introduce \algo, a probabilistic model that can infer a flexible array of mesoscale structure from large-scale hypergraph data. 
In its specification, the model blends structural patterns from stochastic block modeling approaches for traditional networks  \cite{holland1983stochastic,airoldi2008mixed,ball2011efficient} with multilinear tensor decomposition approaches for multiplex networks~ \cite{schein_bayesian_2016,debacco2017community,aguiar_tensor_2023,hood2024ell_0}. 
In particular, nodes have overlapping membership in a set of classes which themselves then exhibit higher-order interactions. Class-level interaction rates are governed by an \textit{affinity multi-tensor} $\Lambdaall$, settings of which determine the type of mesoscale structure present in the hypergraph. 
$\Lambdaall$ is very high-dimensional --- we impose a certain low-rank factorization, which makes the model both identifiable, enabling the recovery of meaningful latent structure, and tractable to estimate. ~\looseness=-1

Before describing $\Lambdaall$ in detail, we define hypergraphs mathematically using tensor notation. 
Formally, a hypergraph can be represented as a \textit{multi-tensor}---i.e., a series of tensors---$\Aall := \{A^{\ms{(2)}}, \dots, A^{\ms{(D)}}\}$---where the $d^{\textrm{th}}$ \textit{adjacency tensor} $\Ad$ stores all observed $d$-order interactions, with $d\!=\!2$ corresponding to pairwise interactions, and $d\!=\!D$ to the maximum observed order.~\looseness=-1 

Each adjacency tensor $\Ad \in \mathbb{N}_0^{N \times \dots \times N}$ is a $d$-way symmetric count tensor with entries 
$\Ad_{\edgetype} := \Ad_{\iidx}$, each of which denotes the number of observed interactions between a particular set of $d$ nodes, where the \textit{multi-index}
$\edgetype := (i_1, \, \dots, i_d) \in \Omegad$, takes values in the set of all possible interactions between $d$ nodes.
For instance, for a legislative bills dataset, $\Ad_{\edgetype}$ could be the number of bills that a given group of legislators $i_1, \dots, i_d$ co-sponsored.  We explicitly model the entries $i_1 < \dots < i_d$ and define the remaining entries by symmetry, such that ${|\Omegad| \!=\! {N \choose d}}$ and the remaining (e.g., diagonal) entries are undefined. ~\looseness=-1

The proposed model builds on previous work~\cite{chodrow2021generative,contisciani2022inference,ruggeri2023community} in assuming that the number of observed higher-order interactions between different sets of nodes are conditionally Poisson-distributed---i.e.:
\begin{equation}
\label{eq:likelihood}
  \mathbb{P}(\Aall \mid \Muall) = \mathop{\mathsmaller{\prod}}_{d=2}^D \mathop{\mathsmaller{\prod}}_{\edgetype \in \Omega^{\ms{(d)}}} \text{Poisson}\left(\Ad_{\edgetype};\, \Mud_{\edgetype}\right),
\end{equation}
where $\Mud_{\edgetype} > 0$ is an interaction rate---i.e., $\mathbb{E}[\Ad_{\edgetype}] \!=\! \Mud_{\edgetype}$. In practice, these rates are unknown and we aim to estimate them. The number of such rates is combinatoric, one for each possible higher-order interaction, and typically only a small fraction of interactions are observed in practice (i.e., $\Aall$ is \textit{sparse}), making estimation difficult.~\looseness=-1

\textbf{Latent classes of nodes.} To handle the prohibitive dimensionality of $\Muall$, we assume it admits the following low-rank factorization:~\looseness=-1
\be \label{eq:mu}
  \Mud_{\e} = \sum_{c_1=1}^C \ldots \sum_{c_d=1}^C \Lambdad_{\cidx} \mathop{\mathsmaller{\prod}}_{r=1}^d \theta_{i_r c_r} . 
\ee
Beyond reducing dimensionality, this factorization yields a particular interpretation wherein $\theta_{i_rc_r} > 0$ represents the rate at which node $i_r$ acts as a member of \textit{class} $c_r$ and $\Lambdad_{\cidx} >0$ represents the rate of $d$-order interactions between the classes $c_1, \dots, c_d$. We can interpret each $C$-length vector $\vm_{i} := (\theta_{i1},\dots,\theta_{iC})$ as representing the overlapping class membership of node $i$, and the set of such vectors as collectively forming the $N\times C$ \textit{node-class membership matrix} $\mathrm{\Theta}$ for a fixed number of classes $C$.~\looseness=-1 

Analogous to the data, the rates $\Lambdad_{\cidx}$ collectively form a multi-tensor $\Lambdaall := \{\mathrm{\Lambda}^{\ms{(2)}},\dots, \mathrm{\Lambda}^{\ms{(D)}}\}$ corresponding to a series of \textit{class affinity tensors}, each of which is $d$-way of size $C \times \dots \times C$. Note that to ensure $\Mud$ is symmetric, $\Lambdad$ should also be symmetric, a condition our parameterization imposes.~\looseness=-1 

Particular settings of $\Lambdaall$ encode different mesoscale patterns regulating hyperedge formation. For example, a purely diagonal multi-tensor where $\Lambdad_{\cidx}$ is only nonzero if $c_1 \!=\! \cdots \!=\! c_d$ encodes strict assortativity. In fact, in this setting the proposed model coincides exactly with the model recently proposed by~\cite{contisciani2022inference}. The proposed model class introduced in this paper is much more general however, allowing for arbitrary higher-order interaction between classes.~\looseness=-1

The parametrization in \Cref{eq:mu} offers a parsimonious and interpretable framework for modeling hypergraphs but still presents significant practical challenges. While it drastically reduces the dimensionality of the original data $\Aall$, the number of parameters in $\Lambdaall$ still grows exponentially in $D$, totaling $\sum_{d=2}^D C^d$ terms, a prohibitively large number for even moderate $C$ and $D$. Moreover, this particular decomposition is generally non-unique~\cite{kolda_tensor_2009}, complicating the interpretation of the estimated parameters.~\looseness=-1

\textbf{Latent communities of classes.} We draw an analogy to the data and view $\Lambdaall$ as itself a \textit{latent hypergraph} of higher-order interactions among classes. We then posit that $\Lambdaall$ admits a low-rank factorization: 
\be\label{eq:core-param1}
\Lambdad_{\cidx} = \sum_{k=1}^{K} \gamma^{\mathsmaller{(d)}}_k \prod_{q=1}^{d} \mathrm{w}_{c_{q} k} ,
\ee

where here we introduce a set of $K$ class-level blocks we call \textit{communities}. The $k^{\textrm{th}}$ community is defined as a distribution over classes given by the column $\mathbf{w}_k$ of a $C \times K$ \textit{class-community membership matrix} $\mathrm{W}$, which is shared across $d \in \{2, \dots, D\}$. This matrix plays a role analogous to the node-class membership matrix $\mathrm{\Theta}$ (see~\Cref{fig:toy}). However, unlike the factorization in~\Cref{eq:mu}, \Cref{eq:core-param1} imposes assortativity on the mesoscale structure of $\Lambdaall$---i.e., that \textit{classes} interact exclusively within (not between) communities. The rate of such interactions in a given community $k$ is then given by $\gamma^{\mathsmaller{(d)}}_k > 0$, which is allowed to vary by the order $d$. We refer to this term as the community-order rate. The $\gamma_{k}^{\ms{(d)}}$ and $\mathrm{W}$ together parameterize $\Lambdaall$, greatly reducing the number of model parameters. This factorization ensures that $\Lambdaall$ is symmetric, which in turn ensures that $\Muall$ is symmetric.~\looseness=-1

\begin{figure*}[!ht]
    \centering
\includegraphics[width=\linewidth]{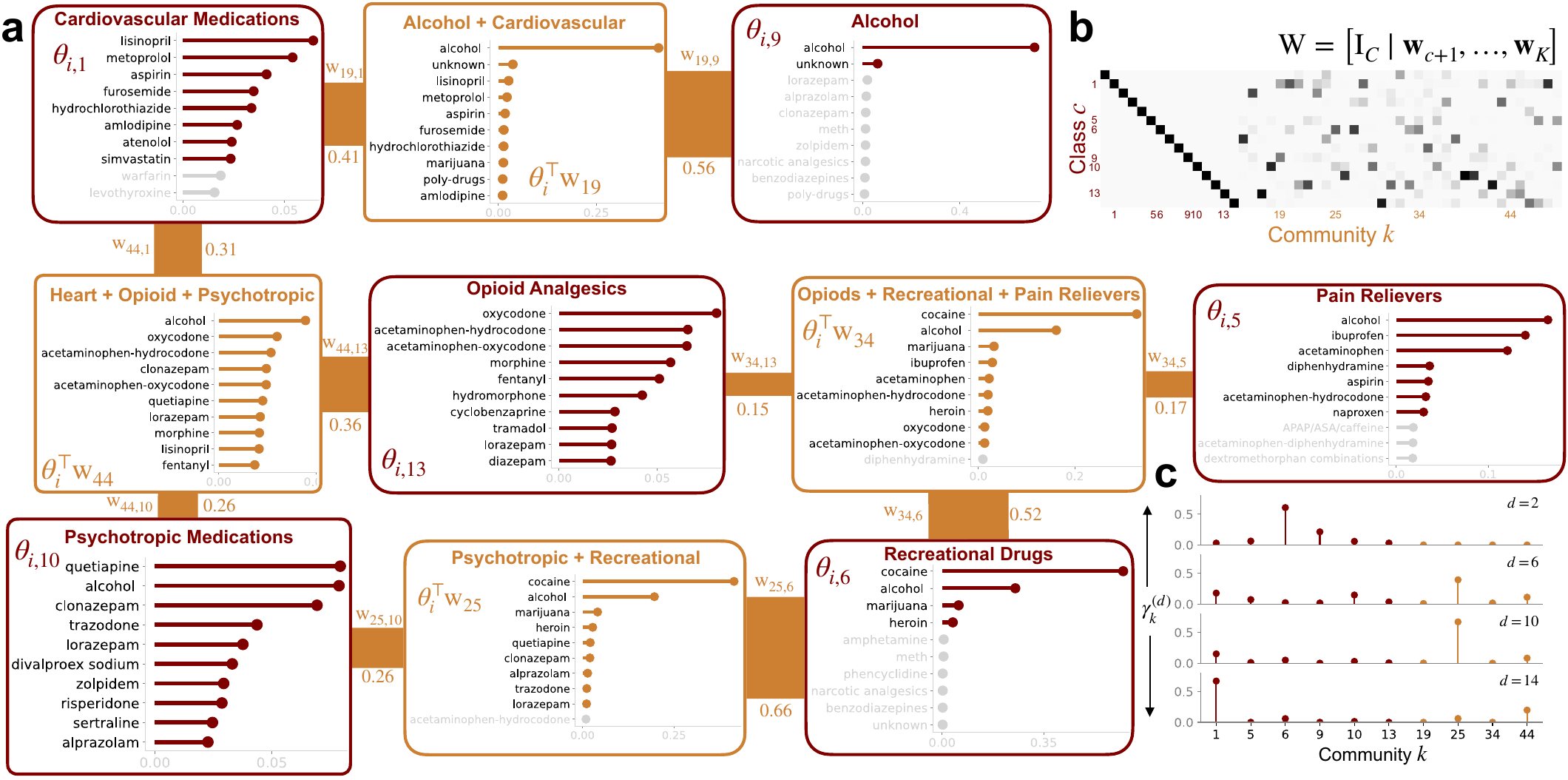}
    \caption{\textbf{\algo~learns a latent hypergraph between identified drug classes in drug-drug interaction data.} a) We 
   highlight six classes of drugs $c \in \ccup{1,5,6,9,10,13}$ (maroon) and four communities, each defined by a convex combination of classes, $k \in \ccup{19,25,34,44}$ (bronze). The columns of the class-community matrix $\mathrm{W}$ (entries correspond to bronze links) define the class weights corresponding to each community. Here, communities may be interpreted as mixtures of classes. The width of the edge between the $c^{\textrm{th}}$ class and $k^{\textrm{th}}$ community is proportional to $\mathrm{w}_{ck}$, representing the weight of class $c$ in community $k$. We show the top drugs occurring in each class, as measured by the node-class membership matrix $\mathrm{\Theta}$, and top drugs in each community, where the node-community loadings are given by the matrix multiplication $\mathrm{\Theta} \mathrm{W}$ (drugs with small nonzero values are grayed). b) $\mathrm{W},$ the class-community matrix. Each element $\mathrm{w}_{ck}$ is the weight of class $c$ in community $k$ and each column sums to 1.  c)  Conditional on the classes and communities shown in (a), for a fixed $d$, shown are the normalized community-order rates of each class and community given by $\gamma^{\mathsmaller{(d)}}_k$. 
   }
    \label{fig:drug}
\end{figure*}

 \begin{figure}[!ht]
    \centering        \includegraphics[width=\linewidth]{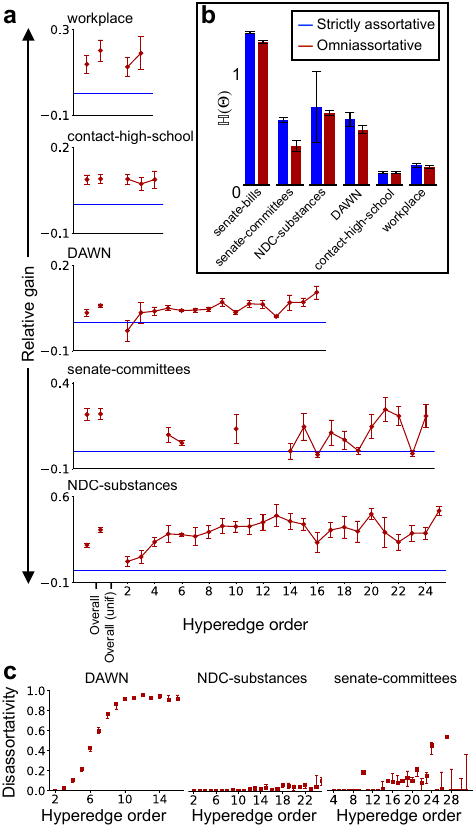}
        \caption{\textbf{Relaxing strict assortativity improves interpretability and link prediction.} a) For each dataset, we show relative gain in heldout log-likelihood  over the strictly assortative baseline. Positive values indicate better link prediction for \algo, with error bars denoting variability over five train-test splits. b) Median entropy of $\boldsymbol{\theta}_i$ across nodes $i$, with errors denoting its interquartile range across five random initializations. Lower values denote more interpretable class structure with less uniform mixed-membership. 
        c) Inferred disassortativity levels vary by hyperedge order and dataset. 
        }
    \label{fig:prediction}
\end{figure}

\vspace{0.5em}
\textbf{Omniassortativity.} The parameterization in \Cref{eq:core-param1} allows nodes of different classes to interact disassortatively, as off-diagonal elements of $\Lambdaall$ can be nonzero. Imposing that interactions among classes occur exclusively within communities mitigates the exponential blowup in parameters without encoding overly restrictive assumptions about the mesoscale structure of the observed hypergraph $\Aall$.~\looseness=-1 

To realize a high degree of disassortativity---i.e., where $\Lambdad$ has much greater off- than on-diagonal elements---some of the parameters $\gamma^{\mathsmaller{(d)}}_k$ must be allowed to be negative. If all are non-negative (and not trivially zero), we show in Supplementary Note 4 that the highest possible degree of disassortativity is bounded below by
\begin{align*}
    \frac{\sum_{c=1}^C \Lambdad_{c \dots c}}{\sum_{c_1=1}^C \ldots \sum_{c_d = 1}^C \Lambdad_{c_1 \dots c_d}} \geq \nicefrac{1}{C^{d-1}}.
\end{align*}
We refer to this sub-family of the model as \textit{semi-assortative}, which allows for departure from strict assortativity, such as in core-periphery structure (and so it is useful in practice), but not to the extreme of strict disassortativity. We show in Supplementary Note 1 how to realize a fully omniassortative model by allowing some $\gamma^{\mathsmaller{(d)}}_k$ to be negative and detail a careful modeling scheme that ensures all elements $\Lambdad_{\cidx}$ remain non-negative. ~\looseness=-1

\vspace{0.5em}
\textbf{Connection to tensor decomposition.} The expression for $\Mud_{\edgetype}$ given in~\Cref{eq:mu} can be equivalently expressed for all entries $\edgetype$ simultaneously as expressing the entire tensor $\Mud$ as following a Tucker decomposition~\cite{tucker_mathematical_1966}:
\begin{equation}
\label{eq:Tucker}
\Mud = \llbracket \Lambdad, \mathrm{\Theta}, \dots, \mathrm{\Theta}\rrbracket
\end{equation}
where, in general, the first argument is the core tensor and the remaining arguments are the factor matrices, one for each mode of the input tensor. In this case, the factor matrices are $d$ repeats of the same node-class membership matrix $\mathrm{\Theta}$, and the core is the class affinity tensor $\Lambdad$. 

Similarly, the expression for $\Lambdad_{c_1 \dots c_d}$ in~\Cref{eq:core-param1} can be equivalently viewed as the tensor $\Lambdad$ following a special case of the Tucker decomposition, called the canonical polyadic (CP) decomposition~\cite{hitchcock1927expression}, where the core tensor is diagonal. See~\Cref{fig:toy}c for a visualization.

\textbf{Model identifiability.} While the Tucker decomposition in~\Cref{eq:mu} is not generally unique, we prove in Supplementary Note 4 that combining~\Cref{eq:core-param1} with two additional assumptions, both of which are easily enforced, renders the model uniquely identified---an important property to ensure reliable parameter estimation. The first assumption is that the columns of $\mathrm{\Theta}$ and $\mathrm{W}$ lie on the probability simplex, that is, their entries are non-negative and sum to $1$. The second is that the first $C$ columns of $\mathrm{W}$ form the identity matrix---i.e.,  
\begin{equation}
\label{eq:W}
[\mathbf{w}_{1}, \dots, \mathbf{w}_{C}] = \textrm{I}_C.
\end{equation}
In enforcing the latter assumption, we impose the number of communities exceeds the number of classes so that ${K \geq C}$. This structure enforces that every class $c$ has a corresponding \textit{pure community} $k\!=\!c$ within which no other classes have membership. In these communities, nodes interact in a strictly assortative manner. ~\looseness=-1 

These assumptions, satisfied in all of our experiments, allow for classes and communities to be identified from the data. This is a nontrivial result that arises from combining uniqueness results from tensor decomposition and non-negative matrix factorization \cite{gillis2020nonnegative,hitchcock1927expression}. For proofs and additional results, see Supplementary Note 4.  ~\looseness=-1

\subsection{Automatic discovery of drug classes and their interactions from large-scale pharmacological data} \label{sec:drugs}
 Here, we apply~\algo~to analyze the Drug Abuse Warning Network (DAWN) database~\cite{benson2018simplicial}. In this setting, nodes are drugs, and each hyperedge represents a set of drugs that an emergency room patient self-reported to having taken. There are 2,558 nodes and 141,178 hyperedges in this dataset; see \Cref{tab:dataset_stats} for additional details. Selecting $C\!=\!15$ classes and $K\!=\!50$ communities (refer to Supplementary Note 3 for details regarding the selection criteria), we fit the model to the entire dataset, and perform an exploratory analysis of the inferred latent structure.~\looseness=-1
 
\pagebreak 

\textbf{Inferred classes of drugs.} We begin by interpreting the inferred classes of drugs, represented by the node-class membership matrix $\mathrm{\Theta}$. In \Cref{fig:drug} we visualize six of the classes as maroon stem plots, where the drugs $i$ with the largest values of $\theta_{ic}$ in each class $c$ are shown. The labels given to each class are assigned by OpenAI's \texttt{GPT-4o} (accessed May 4, 2025), which we prompted with a list of each class's top drugs. We find that the inferred classes often represent groups of drugs with a shared function or purpose, such as ``cardiovascular medications" (e.g.,~lisinopril, metoprolol), ``opioid analgesics" (e.g.,~oxycodone, morphine), or ``psychotropic medications'' (e.g., quetiapine, clonazepam). We note that drugs in the same class are not necessarily drugs that are often found to have been taken together by ER patients. For example, while the drugs in the inferred ``cardiovascular medication'' class have a similar pharmacological function, they typically occur in our data with drugs of other inferred classes---i.e., this is a disassortative class.  With the exception of ``alcohol'', which is present in many classes due to its extreme prevalence in the data, we find that the classes' top drugs are highly coherent and accord with the label assigned to them by \texttt{GPT-4o}. ~\looseness=-1

\textbf{Inferred communities of drugs.} We next inspect the inferred mixtures of classes represented by the class-community membership matrix $\mathrm{W}$. Recall that each column of this matrix $\mathbf{w}_k$ is the distribution over $C$ classes defining community $k$. In~\Cref{fig:drug} we visualize as bronze stem plots four different communities, where we show the top drugs $i$ in each community $k$ which have the largest values of $\boldsymbol{\theta}_i^\top\mathbf{w}_k =  \sum_{c=1}^C \theta_{ic}\mathrm{w}_{ck}$. According to~\Cref{eq:core-param1}, the top drugs in community $k$ should often occur in the data together in $d$-order hyperedges if there is a large value for the parameter $\gamma^{\mathsmaller{(d)}}_k$. We chose to visualize these four communities because, according to $\mathrm{W}$, they were all mainly a mixture of some subset of the six classes we chose to visualize---the bronze bars between classes and communities indicate that mixture, with the width of the bar corresponding to the magnitude of the mixture weight. The community labels are then just combinations of the relevant class labels. 

One of these inferred communities represents a two-way mixture of ``psychotropic medications'' and ``recreational drugs'' and has large weights for $d\!=\! 6$ and $d\!=\!10$, as shown in~\Cref{fig:drug}c. One hypothesis this supports is that some contingent of patients who self-reported taking many (e.g.,~$d \gg 2$) drugs may have suffered a bad interaction between recreational drugs like cocaine and psychotropics like quetiapine. The particular mixture of quetiapine and cocaine---which are the top drugs in both classes, respectively---is actually referred to as a ``Q-ball'' and already known by researchers to cause increased sedation and cardiovascular strain, among other adverse effects~\citep{waters2007intravenous}.~\looseness=-1

Given the structure we impose on $\mathrm{W}$ (see in~\Cref{eq:W}), each class $c$ is associated with a \textit{pure community} $k\!=\!c$ which represents only assortative interactions between nodes of that class. We can see in~\Cref{fig:drug}c that for $d\!=\!2$, a large weight is placed on $k\!=\!6$, which is the pure community for the ``recreational drugs'' class ($c\!=\!6$), suggesting that many ER patients who have only taken two drugs have taken two recreational drugs. We also see that for $d\!=\!14$, a large weight is placed on community $k\!=\!1$ which is the pure community for the ``cardiovascular'' class ($c\!=\!1$), suggesting that many ER patients who have taken 14 drugs, have taken 14 cardiovascular drugs. We surmise this is due to a non-causal correlation in the data---i.e., that patients who are on many cardiovascular medications are those who are already at high risk of heart attack, and wind up in the ER despite (not because of) the drugs.~\looseness=-1 

\begin{figure*}[!ht]
    \centering
    \includegraphics[width=\linewidth]{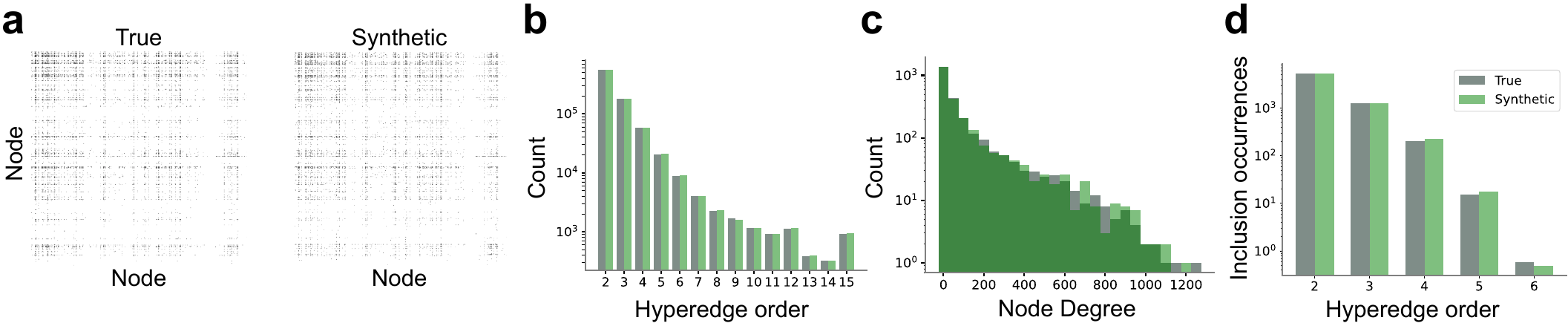}
    \caption{\textbf{Generating synthetic data with \algo.} We use \algo\ to generate data using parameters learned from fitting to the \texttt{DAWN} dataset. We use several metrics to compare the synthetic multi-tensor $\hat{\mathcal{A}}^{\ms{(:)}}$ to the true observed data $\Aall$. a) Projected adjacency matrices of the true (left) and synthetic datasets (right). Hyperedge counts are shown in log scale. b) Number of inclusion occurrences for each hyperedge order $d$ in randomly sampled subsets; this is the number of nonzero hyperedges of size $d$ which appear as a subset of a hyperedge of size $d+1$. c) Empirical node degree distribution. d) Empirical hyperedge order distribution.}
    \label{fig:generation}
\end{figure*}

\subsection{Applicability across domains: prediction, interpretability, and heterogeneity} \label{sec:application}
\textbf{Datasets.} We consider six hypergraphs derived from empirical data: two drug interaction datasets (\texttt{DAWN}, \texttt{NDC-substances})~\cite{benson2018simplicial, chodrow2021generative}, two congress-level datasets (\texttt{senate-committees}~\cite{stewart2017congressional}, \texttt{senate-bills}~\cite{fowler2006legislative, fowler2006connecting, chodrow2021generative}), and two additional human-contact interaction datasets (\texttt{workplace}~\cite{genois2015data}, \texttt{contact-high-school}~\cite{mastrandrea2015contact}). Details on each dataset are provided in the Methods section and \Cref{tab:dataset_stats}. ~\looseness=-1

\textbf{Prediction and model selection.} The proposed model allows for recovery of mesoscale structure by specifying the number of classes ($C$) and communities ($K$). To select the most plausible generative mechanism underlying the data, we compare the heldout likelihoods obtained by the model under various choices of $C$ and $K$ in a link prediction experiment and cross-validate to select the most appropriate $C$ and $K$. Under strict assortativity, the best fit occurs when the number of classes and communities are equal $(K\!=\!C)$. Otherwise, modeling omniassortative structure (setting $K > C$) offers gains in predictive performance.~\looseness=-1

Specifically, for each hyperedge order $d$, we randomly mask either 1,000 of the nonzero counts or 10\% of them, whichever is smaller. We also randomly mask an equal number of zeros. We denote the full set of masked hyperedges by $\Omega_{\textrm{test}}^{\ms{(d)}}$. 
We then fit the model to the unmasked portion of the data.~\looseness=-1

After training, we assess model fit using three metrics based on the log-likelihoods of the masked hyperedges. We employ multiple evaluation metrics because aggregated measures can obscure important differences in model fit. For instance, a model that performs well on pairwise interactions may achieve a similar overall log-likelihood to one that excels at higher-order interactions
$(d \geq 3)$, despite capturing fundamentally different structural aspects of the hypergraph. Therefore, we compute the heldout log-likelihood of each hyperedge $\edgetype$ in the test set $\Omega_{\textrm{test}}^{\ms{(d)}}$ and compute
$$\mathcal{L}^{\ms{(d)}} = \sum_{\edgetype \in \Omega_{\textrm{test}}^{\ms{(d)}}} \log \mathbb{P}(A_{\edgetype}^{\ms{(d)}} \mid \Mud_{\edgetype})$$ 
for each $d$. 
We also compute aggregate metrics
$$\mathcal{L} = \sum_{d=2}^D \mathcal{L}^{\ms{(d)}} \, \text{ and } \mathcal{L}^{(\text{uniform})} = \sum_{d=2}^D \frac{1}{|\Omega_{\textrm{test}}^{\ms{(d)}}|}\mathcal{L}^{\ms{(d)}},$$
which are the heldout log-likelihood, and a weighted heldout log-likelihood which gives uniform weight to each order, respectively. These metrics are shown in \Cref{fig:prediction}a as ``overall" and ``overall (unif)". ~\looseness=-1

For each dataset, we repeat this procedure over a grid of $C$ and $K$, evaluating the combinations for which ${C \leq K}$. The choice of $C$ and $K$ which achieves the highest $\mathcal{L}^{(\text{uniform})}$ value and the corresponding grids are reported in~\Cref{tab:dataset_stats}.~\looseness=-1

\begin{table*}
\centering
\setlength{\tabcolsep}{3pt} 
\begin{tabular}{lrrrrrrrrcc}
    \toprule
     Dataset& $K$ & $C$ & $\%(d\!=\!2)$ & $D$ & $\bar d$ & $\mathcal{A}^{\ms{(\bullet)}}$ & $N_{\textrm{nz}}$ & $N$ & $C$ grid & $K$ grid \\
    \midrule
    \texttt{hospital} (semi-synthetic) & 3 & 2 & 72.5 & 5 & 2.3 & 8216 & 792 & 73 & - & - \\
    \texttt{justice} & 6 & 3 & 0.0 & 9 & 6.4 & 1310 & 282 & 15 & - & - \\
    \texttt{DAWN} & 48 & 16 & 22.4 & 16 & 4.0 &  834,643 & 141,178 & 2,558 & (16, 32, 48, 64) & (16, 32, 48, 64) \\
    \texttt{NDC-substances} & 64 & 32 & 17.9 & 25 & 8.0 & 29,296 & 6,383 & 5,539 & (16, 32, 48, 64, 80)  & (16, 32, 48, 64, 80) \\
    \texttt{workplace} & 8 & 5 & 94.1 & 4 & 2.1 & 9,645 & 788 & 92 & (5, 8, 11, 14) & (5, 8, 11, 14)\\
    \texttt{contact-high-school} & 15 & 12 & 70.3 & 5 & 2.3 & 172,035 & 7,818 & 327 & (9, 12, 15, 18) & (9, 12, 15, 18)\\
    \texttt{senate-bills}   & 24  & 24  & 17.8 & 40 & 8.2 & 29,157 & 21,830 & 294 & (12, 24, 36, 48) & (12, 24, 36, 48)\\
    \texttt{senate-committees} & 6 & 4  & 0.0 & 31 & 17.6 & 315 & 302 & 282 &(2, 4, 6, 8, 10) & (2, 4, 6, 8, 10)\\
    \midrule
\end{tabular}
\caption{\textbf{Dataset summary statistics.} We report the number of communities $K$, classes $C$, the percentage of pairwise interactions $\%(d\!=\!2)$, maximum hyperedge size $D$, mean hyperedge size $\bar d$, sum of hyperedges $\mathcal{A}^{\ms{(\bullet)}}$, number of nonzero hyperedges $N_{\textrm{nz}}$, number of nodes $N$, and the grids used to perform the grid search to select $C$ and $K$.}
\label{tab:dataset_stats}
\end{table*}

The log-likelihoods vary drastically, and we normalize them to compare results across datasets as follows. 
We compute \textit{relative gain} over the strictly assortative baseline model~\cite{contisciani2022inference}; for a given $d$ this is $$\mathcal{RG}^{\ms{(d)}} = \frac{\mathcal{L}_{\textrm{omni}}^{\ms{(d)}} - \mathcal{L}_{\textrm{assort}}^{\ms{(d)}}}{|\mathcal{L}_{\textrm{assort}}^{\ms{(d)}}|},$$
with similar formulas applied to $\mathcal{L}$ and $\mathcal{L}^{(\text{uniform})}$. 

In \Cref{fig:prediction}a, we show the relative gain metrics for the datasets where the omniassortative model performs best (five of the six datasets). That is, the highest values of $\mathcal{L}_{\text{omni}}$ occur when $ K>C$. 
Both overall metrics agree, yielding positive relative gains of varying amounts that range from $0.06$ on the \texttt{DAWN} data to a peak of $0.33$ on the \texttt{NDC-substances} data. The relative gain generally increases with hyperedge order.

Strict assortativity yields the highest prediction performance in \texttt{senate-bills} $(C\!=\!K\!=\!24)$. This result is consistent with previous results obtained for this type of data~\cite{veldt2023combinatorial}, where strong notions of group homophily are found in groups of congress members formed by cosponsorship of legislative bills. This dataset is not shown in \Cref{fig:prediction}a because all values are trivially zero.  ~\looseness=-1

\textbf{Class cohesiveness and interpretability. }
Beyond prediction, we assess whether a more expressive affinity tensor yields more interpretable inferred node-level memberships. 
For each dataset, we compare the proposed model with $(C, K)$ as selected in the prediction task (for \texttt{senate-bills}, since the optimal combination occurs for $C\!=\! K \!=\! 24$, we choose $C \!=\! 24$ and $K \!=\! 36$) to the strictly assortative baseline~\cite{contisciani2022inference}, where $C\!=\!K$. The models share the same choice of $C$. 
We fit each model to the full datasets and compute the empirical entropy of each node's normalized class membership vector $\bar { \bs{\theta}}_i$.
This value is given by ${\mathbb{H}(\bar{ \bs{\theta}}_i) = - \sum_{c=1}^C \bar \theta_{ic} \log(\bar \theta_{ic})}$, which is non-negative. We compute the median value over nodes, reported as $\mathbb{H}(\mathrm{\Theta})$. In the extreme case, each node is assigned to exactly one class and $\mathbb{H}(\mathrm{\Theta}) \!=\! 0$. 
Lower values of $\mathbb{H}(\mathrm{\Theta})$ represent less mixed membership across classes and correspond to more interpretable class structure. ~\looseness=-1

\algo~tends to infer less mixed class memberships than the strictly assortative model, as shown in \Cref{fig:prediction}b. The omniassortative model achieves lower $\mathbb{H}(\mathrm{\Theta})$ values for five of the six datasets (\texttt{senate-bills}, \texttt{senate-committees}, \texttt{NDC-substances}, \texttt{DAWN}, and \texttt{workplace}) while for \texttt{contact-high-school}, the values are similar.~\looseness=-1  

\textbf{Disassortativity varies by interaction order.}
We analyze how disassortativity varies with hyperedge order using \algo~with $C$ and $K$ selected based on the link prediction results. Given the estimated model parameters, we assign hyperedges to class combinations $(c_1, \dots, c_d)$. Interactions are strictly assortative when all classes are equal $( c_1 \!=\! \cdots \!=\! c_d)$ and non-assortative otherwise. For each hyperedge order, we compute the proportion of hyperedges involving multiple classes. We use this proportion as a measure of disassortativity; see the Methods section for further details regarding how it is defined and computed.~\looseness=-1

\Cref{fig:prediction}c shows these proportions for three higher-order datasets. Each dataset contains higher-order interactions with maximum order ($D$) exceeding 15. In each setting, modeling omniassortativity improves prediction performance. 
The qualitative behavior of the curves varies by dataset. For example, in \texttt{DAWN} we observe a monotonic increase in disassortativity until about order $d\!=\!10$, after which the proportion plateaus near $1$. These results illustrate the importance of carefully modeling omniassortativity in the hypergraph setting to appropriately capture underlying latent structure.~\looseness=-1

\subsection{Fast hypergraph generation}
Finally, \algo~is a generative model, and here we show how to use it to generate synthetic hypergraphs with prespecified mesoscale structures. 
The problem of hypergraph generation remains an open problem due to computational challenges~\cite{ruggeri2024framework,kaminski2023hypergraph,chodrow2021generative}. 
However, exploiting properties of the proposed model allows us to address this task tractably. We describe an algorithm in the Methods section which generates hypergraphs of arbitrarily-sized orders and whose computational complexity is linear in the expected number of hyperedges. ~\looseness=-1

To illustrate this algorithm empirically, we generate synthetic hypergraph data using the model parameters learned on the \texttt{DAWN} data, as described in~\Cref{sec:drugs}. Our method is fast: running on a personal laptop with one CPU, in just 70 seconds it generates 833,564 higher-order drug interactions (similar in number to the true data) ranging from order $d=2$ to $d=16$.~\looseness=-1

To check whether the synthetic data is similar to the true data, we compare a number of statistics: the node-degree distributions, hyperedge order distributions, inclusion occurrence distributions~\cite{lotito2022higher}, and projected adjacency matrices. 
\Cref{fig:generation} shows these comparisons. 
The plots appear nearly identical, demonstrating that the synthetic data closely resembles the true data. ~\looseness=-1

\section{Conclusions}
This paper introduces a family of probabilistic generative models for higher-order interaction data that can represent and efficiently discover a broad spectrum of mesoscale structure in large-scale hypergraphs, from strictly assortative to disassortative. Different kinds of such structure offer fundamentally different descriptions of complex systems. It is thus critically important for understanding and predicting the behavior of such systems to be able to flexibly model a broad spectrum of structure in their underlying hypergraph. ~\looseness=-1

We resolve the central computational challenge presented by higher-order interaction data by combining block modeling techniques with low-rank factorized representations. The proposed model blocks nodes into classes and then represents higher-order interactions among classes. The key insight is to view classes as themselves forming a latent hypergraph whose structure is strictly assortative. Unlike prior work which assumes the observed hypergraph is assortative, this assumption still permits the observed graph to take on a broad range of structure, as we prove. The proposed model thus exploits the computational benefits of assortative structure among latent classes while still being able to represent and discover omniassortative structure among observed nodes.~\looseness=-1  

Our theoretical results provide rigorous guarantees that guide and license the model's use in a wide range of settings. Among them, we show that the model respects the symmetry of the data, that it generalizes existing strictly assortative models, and that it is identifiable.~\looseness=-1

Empirically, we demonstrated that the model is sufficiently flexible to capture a variety of latent mesoscale structure in a variety of higher-order network datasets arising from various real-world social, political, and biomedical settings. 
We illustrated how mesoscale structure can vary with hyperedge order, as the same class of nodes may behave assortatively in hyperedges of one order but disassortatively in another. Through a case study of drugs taken by patients in emergency hospital visits, we have demonstrated how the proposed model is able to learn classes drugs with similar function, and to predict which drugs of different classes are likely to have been taken together by ER patients. Comparisons between modeling approaches in two more case studies, one within hospitals and another concerning Supreme Court Justices, underscore the proposed method's significance.~\looseness=-1 

Finally, we have shown how to efficiently sample synthetic hypergraphs with a pre-specified mesoscale structure under the proposed framework. We have presented an algorithm, derived from the model's probabilistic properties, that is capable of rapidly producing large-scale hypergraphs exhibiting diverse structures and varying orders. We have empirically demonstrated the algorithm's speed by quickly generating large hypergraphs at scale. Moreover, we have shown that the synthetic data closely matches real-world hypergraph data. This result has far-reaching impact, potentially enabling researchers to study realistic higher-order networks of various structures in controlled settings. ~\looseness=-1

Our results point towards many directions for future work. The proposed model allows hyperedges of different orders to contribute differently to community structure. In several cases however, hyperedges are observed to express nestedness or other forms of hierarchy \cite{mariani2019nestedness,landry2024simpliciality,larock2023encapsulation,joslyn2020hypernetwork}. These may create strong correlations between hyperedges of different orders that may call for imposing explicit functional dependencies between these parameters, e.g. between parameters corresponding to hyperedges of subsequent orders. We considered here hypergraphs with discrete weights, but a natural model extension would be to allow for real-valued weights, such as through a compound Poisson construction~\cite{zhou2016augmentable,basbug2016hierarchical}.
Similarly, we focused here on static hypergraphs, where the set of nodes and hyperedges is fixed. In temporal hypergraphs, where the structure varies in time, it would be more appropriate to incorporate dynamical mechanisms for hyperedge formations directly into the model \cite{gallo2024higher,iacopini2024temporal,chowdhary2021simplicial,he2025hypergraph,benson2018sequences}.~\looseness=-1
In the presence of node attributes in hypergraphs, beyond the information contained intrinsically in hyperedges, 
a natural extension would be to incorporate this extra information into the model formulation \cite{badalyan2024structure,contisciani2025flexible}. Finally, the proposed model targets undirected hypergraphs, where the order of nodes organized in hyperedges is not defined. A future direction would be to adapt our formulations to directed hypergraphs~\cite{gallo1993directed}, where one can identify subset of nodes that play different roles, e.g. sender and receiver nodes in a network. In this setting, some symmetries break down, potentially allowing for the incorporation of ideas from asymmetric tensor decompositions for modeling multi-layer network data ~\cite{debacco2017community, aguiar_tensor_2023, hood2024ell_0}.~\looseness=-1

\vspace{1em}
\textbf{Acknowledgements:}
JH is supported by the National Science Foundation under Grant No. 2140001. 
The authors declare no competing interests.

\textbf{Data and code availability:}
The datasets used in the paper are publicly available from their sources listed in the Methods section and in the Supplementary Materials. The source code for the model's algorithmic implementation is open-source and publicly accessible at \url{https://github.com/jhood3/Hypergraphs}. ~\looseness=-1

\newpage

\bibliographystyle{ScienceAdvances}
\bibliography{bibliography}

\newpage 

\section{Methods}
\subsection{Parameter estimation}

To estimate the parameters $\mathrm{\Theta},\mathrm{W}$, and $\mathrm{\Gamma} := \left(\gamma_{k}^{\ms{(d)}}\right)_{d, k}$, given the input data $\Aall$, we perform maximum likelihood estimation. Specifically, we maximize the log-likelihood, proportional in $(\mathrm{\Gamma}, \mathrm{W}, \mathrm{\Theta})$ to
 \begin{align}\label{eq:log-lik}
      \mathcal{L}(\Aall, \Muall) = - \sum_{d=2}^D \sum_{\edgetype \in \Omega^{\ms{(d)}}} \Mud_{\edgetype} + \sum_{A_{\edgetype}^{\ms{(d)}} > 0} A_{\edgetype}^{\ms{(d)}} \log(\Mud_{\edgetype}) ,
 \end{align}
 where $\Omega^{\ms{(d)}}$ is the set of all size-$d$ combinations between $N$ nodes such that $|\Omega^{\ms{(d)}}| \!=\! {N \choose d}$.~\looseness=-1

We derive a generalized expectation-maximization (EM) algorithm~\cite{dempster1977maximum} specific to the proposed model which exploits the sparsity in the data and the low-rank, shared parameterization of the affinity tensors to quickly estimate parameters. The algorithm relies on the Poisson latent subcount representation of the observed data~\cite{schein2019allocative, yildirim2020bayesian}. Under this representation, each observed count $A_{\edgetype}^{\ms{(d)}}$ is the sum of $C^d \cdot K$ conditionally independent latent counts indexed by multi-index $\bs{c} = (c_1, \dots, c_d)$ and community index $k \in [K]$, given by
\be\label{eqn:latcounts}
A_{\edgetype \bs{c} k}^{\ms{(d)}} \overset{\text{ind.}}{\sim} \text{Poisson}(\gamma^{\mathsmaller{(d)}}_k \mathop{\mathsmaller{\prod}}_{r=1}^d \mathrm{w}_{c_rk}\,\m_{i_r c_r}).
\ee
\hspace{1em}
We model these latent counts to construct and maximize an evidence lower bound to the log-likelihood until the evidence lower bound converges or a stopping criterion is reached. The fixed point is guaranteed to be a local maximum, but not the global maximum. 
Therefore, we perform ten runs of the algorithm with different random initializations for $(\mathrm{\Gamma}, \mathrm{W}, \mathrm{\Theta})$, and take the fixed point with the highest
log-likelihood as in \Cref{eq:log-lik}.  We discuss the generalized EM algorithm in greater detail in Supplementary Note 1, where we give exact updates and derivations for each of the strictly assortative, semi-assortative, and omniassortative models. ~\looseness=-1

Alternatively, we may also place prior distributions on the parameters and extend to a maximum a posteriori inference procedure. While we do not do this in practice, we provide details in Supplementary Note 2.

\textbf{Computational complexity.} The complexity of parameter estimation varies with parameterization. The tradeoff between model expressivity and complexity is reported in \Cref{tab:complexity}.
The dominant term varies between $O(CN_{\textrm{nz}})$ and $O(CKN_{\textrm{nz}})$, where $N_{\textrm{nz}}=\sum_{d=2}^D ||\Ad||_0$ is the number of nonzero entries of $\Aall$, making the model scalable and applicable in practice. 
 The fastest and least expressive model is the strictly assortative model. The most expressive---albeit most expensive---model is the omniassortative model, which includes an additional factor of $K$ compared to the faster assortative version.
 The semi-assortative model offers a middle ground. We fit the semi-assortative model in all of our experiments except for those performed on the semi-synthetic \texttt{hospital} data. ~\looseness=-1
 
\begin{table}[!htbp]
    \centering
\resizebox{\columnwidth}{!}{
    \begin{tabular}{>{\centering\arraybackslash}p{1.1cm} >{\centering\arraybackslash}p{0.7cm} >{\centering\arraybackslash}p{1.2cm} >{\centering\arraybackslash}p{1.3cm}}
    \toprule
         &\makecell{Strictly \\ assortative}& \makecell{Semi-\\ assortative} & \makecell{\hspace{-0.3cm}Omni-\\ assortative} \\
      \midrule
      E step  & O$(C N_{\textrm{nz}})$  & O$(K N_{\textrm{nz}})$ & O$({KC}N_{\textrm{nz}})$  \\
      M step & O$(NDC)$ & O$(NDCK)$ & O$(NDCK)$\\ \midrule
      Assortativity & \textcolor{green}{\ding{51}} & \textcolor{green}{\ding{51}} & \textcolor{green}{\ding{51}} \\
      \smash{Core-periphery} & \textcolor{red}{\ding{55}} & \textcolor{green}{\ding{51}} & \textcolor{green}{\ding{51}} \\ 
      Disassortativity & \textcolor{red}{\ding{55}} & \textcolor{red}{\ding{55}} & \textcolor{green}{\ding{51}} \\
      \bottomrule
    \end{tabular}}
    \caption{\textbf{Tradeoff between expressivity and computational complexity.} Here $N_{\textrm{nz}}:=\sum_{d=2}^D ||\Ad||_0$ is the number of nonzero entries of the input set of hypergraph adjacency tensors $\Aall$. }
    \label{tab:complexity}
\end{table}

\subsection{Evaluation metrics}
We consider several metrics to evaluate the proposed method. Here we provide more details describing and defining these metrics.~\looseness=-1

\textbf{Latent subcounts.} Several of the metrics in this paper are functions of latent subcounts~\cite{schein2019allocative, yildirim2020bayesian}. Starting from the multi-index subcounts in \Cref{eqn:latcounts}, we can either consider latent community subcounts:
\be\label{eqn:comsub}
A_{\edgetype k}^{\ms{(d)}} :=\sum_{c_1=1}^{C}\ldots \sum_{c_d =1}^C A_{\edgetype\bs{c}k}^{\ms{(d)}}
\ee
or latent multi-class subcounts:
\be\label{eqn:clasub}
A_{\edgetype \bs{c}}^{\ms{(d)}} := \sum_{k=1}^{K}A_{\edgetype\bs{c}k}^{\ms{(d)}}.
\ee

\textbf{Class affinity matrix.}
To interpret the mesoscale structure among classes, we compute
the $C \times C$ class affinity matrix as:
\be\label{eqn:claaffinity}
\sum_{d=2}^D \sum_{k=1}^K \gamma^{\mathsmaller{(d)}}_k \mathbf{w}_k \mathbf{w}_k^\top.
\ee
This is an aggregate measure of affinity across communities and hyperedge orders. It is shown in \Cref{fig:prediction}a and \Cref{fig:prediction}b beneath the node-class memberships $\Theta$.~\looseness=-1

\textbf{Mixed membership.} To measure the mixed-membership levels learned by a model, we consider each node's class membership vector given by $\bs{\theta}_{i} \in \mathbb{R}^C$. We first normalize each node's class membership vector, dividing by $\sum_{c=1}^C\theta_{ic}$ to make it a valid discrete probability distribution over classes and compute its entropy. ~\looseness=-1

For a vector $\mathbf{v}$, the entropy is defined as 
\begin{align*}
    \mathbb{H}(\mathbf{v}) = -\sum_{i}\mathrm{v}_i \log(\mathrm{v}_i),
\end{align*}
where $\mathrm{v}_i \log(\mathrm{v}_i) = 0$ if $\mathrm{v}_i \!=\! 0$. 
Higher entropy indicates greater mixing of a node’s class membership across different classes, which can hinder interpretability. In contrast, the extreme case of zero entropy, $\mathbb{H}(\mathbf{v}) = 0$, corresponds to hard membership, where each node belongs to exactly one class.
 ~\looseness=-1

\textbf{Jensen-Shannon divergence.} We quantitatively measure the dissimilarity of a community indexed by $k$ from an assortative class indexed by $c$ using the Jensen-Shannon divergence between distributions, given by
\begin{align*}
    \text{JS}(\bs{\theta}_c \mid \mid \mathrm{\Theta} \mathbf{w}_k) 
    &= \frac{1}{2}\textrm{KL}( \bs{\theta}_c \mid \mid \bs{M}^{\ms{(c,k)}}) \\
    &\quad + \frac{1}{2} \textrm{KL}(\mathrm{\Theta} \mathbf{w}_k \mid \mid \bs{M}^{\ms{(c,k)}}),
\end{align*}
where $\bs{M}^{\ms{(c,k)}}$ is the mixture of $\bs{\theta}_c$ and $\mathrm{\Theta} \mathbf{w}_k$, given by $\bs{M}^{\ms{(c,k)}} := \frac{1}{2} (\bs{\theta}_c + \mathrm{\Theta} \mathbf{w}_k)$. The Jensen-Shannon divergence is a symmetric alternative to the Kullback-Leibler (KL) divergence~\cite{kullback1951information}. The lower its value, the more similar the two distributions are. For each community $k$, we compute the Jensen-Shannon divergence with each class $c$ and select the minimum ${\min_{c} \text{JS}(\bs{\theta}_c \mid \mid \mathrm{\Theta} \mathbf{w}_k)}$ of these values.  We consider this minimum value to be a measure of non-assortativity. For instance, if $\mathrm{w}_{ck} \!=\! 1$ for a class $c$ and $0$ for all other classes, then $\mathrm{\Theta} \mathbf{w}_k \!=\! \bs{\theta}_{c}$ and the proposed metric is equal to zero.~\looseness=-1 

\textbf{Allocation.} We also compute the expected number of hyperedges allocated to the $k^{\text{th}}$ community, given by $\sum_{d=2}^D \sum_{\edgetype \in \Omegad} A_{\edgetype k}^{\ms{(d)}}$, where we use latent community subcounts as in \Cref{eqn:comsub}. The summation over the space of possible hyperedges $\bigcup_{d=2}^D \Omegad$ can be done efficiently, as we only need to sum over the nonzero entries of $\Aall$, which is linear in the number of observed hyperedges and usually linear in the number of nodes. 

We use these the two metrics described above to identify communities dissimilar from each class and which account for a meaningful portion of the observed hypergraph data (high allocation of hyperedges). In particular, we identify communities $k\!=\!19$ and $k\!=\!44$ as starting points for the exploratory analysis in \Cref{sec:drugs}.~\looseness=-1

\textbf{Proportion of disassortativity.} For each hyperedge order $d$, we compute the expected proportion of hyperedges that are disassortative. Again, we rely on the latent subcount representation. Given latent subcounts $A_{\edgetype \cidx}$, we compute
\begin{align*}
    1 - \frac{\sum_{\edgetype \in \Omegad} \sum_{c=1}^C  A_{\edgetype c \dots c}^{\ms{(d)}}}{\sum_{\edgetype \in \Omegad} A_{\edgetype}^{\ms{(d)}}},
\end{align*}
where the numerator of the second term sums over assortative communities only $(c, \dots,c)$. In practice, we cannot compute this term naively. Doing so requires summing over a combinatorial number of terms. Instead, we leverage the Poisson thinning property.
In particular, for a hyperedge $\edgetype$ and weight $A_{\edgetype}$, we first distribute the weight across communities, as given by $A_{\edgetype k}$. Then, for each community $k$ and hyperedge $\edgetype$ we compute its disassortativity proportion, given by
\begin{align*}
   \rho_{\edgetype k}^{(\text{dis})} =  1 - \frac{\sum_{c=1}^C \mathrm{w}_{ck}^d \prod_{i \in \edgetype}\theta_{ic}}{\prod_{i \in \edgetype} \mathbf{w}_k^\top \bs{\theta}_{i}},
\end{align*}
and multiply by $A_{\edgetype k}$. We then add over all hyperedges and communities and divide by the sum of the hyperedges to reach the aggregate measure of disassortativity proportion given above:
\begin{align*}
\f{\sum_{\edgetype, k} A_{\edgetype k}^{\ms{(d)}}\, \rho_{\edgetype k}^{\text{(dis)}}}{\sum_{\edgetype}\Ad_{\edgetype}} .
\end{align*}

\textbf{AUC.} In the semi-synthetic hospital experiment, we  measure the area under the receiver-operator characteristic curve (AUC) on the heldout data to evaluate model performance. The binary categorization is whether each heldout hyperedge is greater than zero, or zero. The AUC is equal to the probability the model assigns more probability to a randomly selected true positive hyperedge being positive than a randomly selected zero weighted hyperedge~\cite{fawcett2006introduction}. The term 
$$\mathbb{P}(\Ad_{\edgetype} > 0) = 1 - e^{-\Mud_{\edgetype}}$$ 
is increasing in $\Mud_{\edgetype}$, and so we compare rate parameters between hyperedges. That is, for the balanced heldout set $\Omegad_{\textrm{test}}$ consisting of an equal number of zeros and nonzeros, let the nonzeros be denoted by ${\Omega_{\textrm{test}, 1}^{\ms{(d)}}}$ and the zeros be denoted by ${\Omega_{\textrm{test}, 0}^{\ms{(d)}}}$. Then we compute the AUC by randomly pairing zero and nonzero hyperedges. For a given pair $p$ in the set of pairs $\mathcal{P}$, we compute the AUC:
\begin{align*}
   \text{AUC} = \frac{1}{|\mathcal{P}|}\left[\sum_{p}1(\mu_{p,1} > \mu_{p,0}) + 0.5 \sum_{p}1(\mu_{p,0} \!=\! \mu_{p,1})\right].
\end{align*}
$\mu_{p,1}$ and $\mu_{p,0}$ are the rates of the positive and zero-weighted hyperedges in pair $p$, respectively.~\looseness=-1  

\subsection{Hypergraph generation}
The aggregate hyperedge count 
\begin{align*}
\mathcal{A}^{\ms{(\bullet)}} = \sum_{d=2}^D \sum_{\edgetype \in \Omega^{\ms{(d)}}} A_{\edgetype}^{\ms{(d)}}
\end{align*}
is marginally Poisson under~\algo:
\begin{align}
    \mathcal{A}^{\ms{(\bullet)}} \sim \text{Poisson}(\Capmu^{\ms{(\bullet)}}), \, \Capmu^{\ms{(\bullet)}} :=\sum_{d=2}^D \sum_{k=1}^K \gamma^{\mathsmaller{(d)}}_k \phi_{k}^{\mathsmaller{(d)}},
\end{align}
where $\phi_{k}^{\mathsmaller{(d)}} := \sum_{\mathbf{i} \in \Omegad} \prod_{i \in \mathbf{i}} \mathbf{w}_{k}^\top \bs{\theta}_{i}$.
 To sample a hypergraph, we first sample $\mathcal{A}^{\ms{(\bullet)}} $ and thin $\mathcal{A}^{\ms{(\bullet)}} $ over $d$ and $k$ into community counts $A^{\ms{(d,k)}}$. For each $d$ and $k$, we sample $A^{\ms{(d,k)}}$ independent and identically distributed ``hyperevents" by sampling $d$ nodes $(i_1, \dots, i_d)$ without replacement from the set of nodes $[N]$, with sampling weight of node $i$ proportional to $m_{ik}:=\mathbf{w}_k^\top \bs{\theta}_i$. We denote the $i^{\text{th}}$ hyperevent of order $d$ and community $k$ by $\mathbf{e}_i^{\ms{(d,k)}}$. Each hyperedge is then defined as ${\Ad_{\edgetype} := \sum_{k=1}^K \sum_{i=1}^{A^{\ms{(d,k)}}} 1(\mathbf{e}_i\!=\edgetype)^{\ms{(d,k)}}}$.
 Under this procedure, computational cost scales linearly in the count $\mathcal{A}^{\ms{(\bullet)}}$ and rate $\Capmu^{\ms{(\bullet)}}$.~\looseness=-1

\subsection{Description of the datasets}
We consider eight datasets in our experiments, as detailed in \Cref{tab:dataset_stats}.
Each dataset consists of a list of hyperedge occurrences, (which we bin to create count-valued ``weights") and node labels.~\looseness=-1

We analyze three datasets (\texttt{workplace}, \texttt{contact-high-school}, \texttt{hospital}) collected by the SocioPatterns collaboration (\href{http://www.sociopatterns.org}{http://www.sociopatterns.org}). Each dataset describes the interactions between humans in close physical proximity, obtained by wearable sensors. The \texttt{workplace} dataset contains hyperedges that are contacts of employees across five different departments in an office building in France. The \texttt{contact-high-school} dataset describes interactions between students across nine classrooms. The \texttt{hospital} data describes interactions between and among patients and healthcare workers. The semi-synthetic hypergraph constructed from this data removes interactions that are exclusively among healthcare workers.~\looseness=-1 

The following four datasets were constructed in \href{https://www.cs.cornell.edu/~arb/data/}{https://www.cs.cornell.edu/~arb/data/}. The first two concern interactions between substances. The Drug Abuse Warning Network (\texttt{DAWN}) data consists of interactions between drugs, where each interaction is the set of drugs in a patient (as reported by the patient) who visited the emergency department of one of multiple hospitals in the United States. The \texttt{NDC-substances} data consists of drugs found in the National Drug Code directory, where a hyperedge is a drug code consisting of the substances (nodes) that make up that drug.~\looseness=-1 

The third and fourth datasets are United States Senate datasets, \texttt{senate-bills} and \texttt{senate-committees}. In each dataset, nodes are senators. In the bill co-sponsorship data, hyperedges are the set of senators who co-sponsor a bill in the Senate. We restrict our analysis to hyperedges of size $d \leq D \!=\! 40$. In the \texttt{senate-committees} data, hyperedges are the sets of senators serving as members of the same committee. ~\looseness=-1

The \texttt{justice} hypergraph example is constructed from data found in \href{http://scdb.wustl.edu/about.php}{http://scdb.wustl.edu/about.php}. The dataset consists of United States Supreme Court cases and each United States Supreme Court Justice's vote (for or against). Each node is a Justice and each hyperedge corresponds to the set of Justices that sided with the majority opinion of a case.  We consider cases with decisions from 2005-2024, coinciding with the time John Roberts served as Chief Justice up through the beginning of this study.~\looseness=-1 

\clearpage

\beginsupplement
\begin{widetext}

\section*{{Supplementary Materials}}

\section*{Supplementary Note 1: Inference}

\subsection*{Background on the EM algorithm}\label{sec:EM}
In maximum likelihood estimation, the goal is to find the parameter $\bs{\theta}$ that maximizes the likelihood function $p_{\boldsymbol{\theta}}(\mathbf{x})$ of the observed data $\mathbf{x}$ given $\bs{\theta}$. This entails solving
\begin{align}\label{eq:argmax}
\widehat{\bs{\theta}} = \arg \max_{\bs{\theta}}p_{\bs{\theta}}(\mathbf{x}),
\end{align}
or equivalently, maximizing the log-likelihood
\begin{align*}
    \mathcal{L}(\mathbf{x}, \bs{\theta}) = \log p_{\bs{\theta}}(\mathbf{x}). 
\end{align*}

Often, the solution to~\Cref{eq:argmax} is intractable. Starting at an initial point, the expectation-maximization (EM) algorithm~\cite{dempster1977maximum} iteratively updates parameters according to a set of update rules specific to the problem at hand. 

The EM algorithm is particularly useful when introducing a set of latent variables, denoted by $\mathbf{z}$, makes maximizing the complete log-likelihood $\log p_{\bs{\theta}}(\mathbf{x}, \mathbf{z})$ tractable. In this setting, the observed data likelihood is  
\[
p_{\bs{\theta}}(\mathbf{x}) = \int p_{\bs{\theta}}(\mathbf{x}, \mathbf{z}) \, d\mathbf{z}.
\]  

We cannot maximize the complete log-likelihood (since $\mathbf{z}$ is unobserved), but we can maximize its expectation ${\mathcal{Q}(\mathbf{x}, {\bs{\theta}}) = \Exp_{\mathbf{z}}[\log p_{\bs{\theta}}(\mathbf{x}, \mathbf{z}) \mid \mathbf{x}]}$. $\mathcal{Q}(\mathbf{x}, {\bs{\theta}})$ is the \textit{evidence lower bound}, a lower bound on the log-likelihood:~\looseness=-1
\begin{align*}
    \log p_{\bs{\theta}}(\mathbf{x}) 
    &= \log \int p_{\bs{\theta}}(\mathbf{x}, \mathbf{z}) \, d\mathbf{z} \\
    &= \log \int p_{\bs{\theta}}(\mathbf{x}, \mathbf{z}) \frac{ p_{\bs{\theta}}(\mathbf{z} \mid \mathbf{x})}{p_{\bs{\theta}}(\mathbf{z} \mid \mathbf{x})} \, d\mathbf{z} \\
    &= \log \int p_{\bs{\theta}}(\mathbf{z} \mid \mathbf{x}) \frac{p_{\bs{\theta}}(\mathbf{x}, \mathbf{z}) }{p_{\bs{\theta}}(\mathbf{z} \mid \mathbf{x})} \, d\mathbf{z} \\
    &= \log \Exp_{\mathbf{z} \sim p_{\bs{\theta}}(\cdot \mid \mathbf{x})} 
    \left[\frac{p_{\bs{\theta}}(\mathbf{x}, \mathbf{z})}{p_{\bs{\theta}}(\mathbf{z} \mid \mathbf{x})}\right] \\
    &\geq \Exp_{\mathbf{z} \sim p_{\bs{\theta}}(\cdot \mid \mathbf{x})} 
    \left[\log \frac{p_{\bs{\theta}}(\mathbf{x}, \mathbf{z})}{p_{\bs{\theta}}(\mathbf{z} \mid \mathbf{x})}\right]  
    && \text{(by Jensen's inequality)} \\
    &= \Exp_{\mathbf{z} \sim p_{\bs{\theta}}(\cdot \mid \mathbf{x})} 
    \left[\log p_{\bs{\theta}}(\mathbf{x}, \mathbf{z})] - \Exp_{\mathbf{z} \sim p_{\bs{\theta}}(\cdot \mid \mathbf{x})}[\log p_{\bs{\theta}}(\mathbf{z} \mid \mathbf{x}) \right] && \text{(decomposition into $\mathcal{Q}$ and entropy term)} \\
    &= \mathcal{Q}(\mathbf{x}, {\bs{\theta}}) + \mathbb{H}(p(\mathbf{z} \mid \mathbf{x}))
    \\
    &\geq \mathcal{Q}(\mathbf{x}, {\bs{\theta}}). &&  (\mathbb{H}(p(\mathbf{z} \mid \mathbf{x})) \geq 0)
\end{align*}

Often, $p_{\bs{\theta}}(\mathbf{x}, \mathbf{z})$ is easier to maximize (with respect to $\bs{\theta}$) than $p_{\bs{\theta}}(\mathbf{x})$. Starting at initial point $\bs{\theta}_0$, the EM algorithm alternates between computing the expectation (E) step $\mathcal{Q}(\mathbf{x}, \bs{\theta}_t)$ and the maximization (M) step $\bs{\theta}_{t+1} = \arg \max_{\bs{\theta} \in \mathrm{\Theta}} \mathcal{Q}(\mathbf{x}, \bs{\theta})$.
 The log-likelihood $\mathcal{L}(\mathbf{x}, {\bs{\theta}}_t)$ increases at each iteration $t$ of the algorithm~\cite{dempster1977maximum} and the sequence $(\mathcal{L}(\mathbf{x}, \bs{\theta}_t))_{t=0}^\infty$ converges to a local maxima~\cite{wu1983convergence}.\\

 When closed-form updates to $\bs{\theta}$ are not available, the \textit{generalized} EM algorithm naturally extends the EM algorithm. Instead of finding the exact maximizer $\bs{\theta}_{t+1} = \arg \max_{\bs{\theta} \in \mathrm{\Theta}} \mathcal{Q}(\mathbf{x}, \bs{\theta})$, the generalized EM algorithm requires only that the updated parameter increases the expected log-likelihood, i.e., $\mathcal{Q}(\mathbf{x}, \bs{\theta}_{t+1}) > \mathcal{Q}(\mathbf{x}, \bs{\theta}_t)$, at each iteration. The sequence $\{\mathcal{L}(\mathbf{x}, \bs{\theta}_t)\}_{t=0}^\infty$ is still guaranteed to converge to a local maximum. 

\subsection*{Inference: the general pipeline}

We derive a generalized EM algorithm~\citep{dempster1977maximum} to maximize each model's log-likelihood, which is proportional in $(\mathrm{\Gamma}, \mathrm{W}, \mathrm{\Theta})$ to 
 \begin{align}\label{eq:llk}
      \mathcal{L}(\Aall, \Muall) = - \sum_{d=2}^D \sum_{\mathbf{i} \in \Omegad} \Mud_{\mathbf{i}} + \sum_{\Ad_{\mathbf{i}} > 0} \Ad_{\mathbf{i}} \log(\Mud_{\mathbf{i}}).
 \end{align}

We define the complete likelihood by introducing latent subcounts $\Ad_{\mathbf{i} \bs{c} k}$ for each hyperedge $\e = (i_1, \dots, i_d)$, combination of classes $\mathbf{c} = (c_1, \dots, c_d)$ and community $k \in [K]$:
\begin{align*}
     \Ad_{\mathbf{i} \bs{c} k} &\overset{\text{ind.}}{\sim} \text{Poisson}(\gamma^{\ms{(d)}}_k \sprod_{r=1}^d\mathrm{w}_{c_r k} \sprod_{r=1}^d \m_{i_r c_r}),\hspace{1em}
     \Ad_{\mathbf{i}}  = \sum_{c_1=1}^C \ldots \sum_{c_d = 1}^C \sum_{k=1}^K \Ad_{\mathbf{i} \bs{c} k} .
 \end{align*}
 
 The evidence lower bound is proportional in $(\mathrm{\Gamma}, \mathrm{W}, \mathrm{\Theta})$ to~\looseness=-1
\begin{align} \label{eq:elbo_prop}
 \mathcal{Q}(\Aall, \Muall)= 
 -\sum_{d=2}^D \sum_{\mathbf{i} \in \Omegad} [\Mud_{\mathbf{i}} 
     + \sum_{c_1=1}^C \ldots \sum_{c_d = 1}^C \sum_{k=1}^K \Exp \left[\Ad_{\mathbf{i} \bs{c} k} \mid \Ad_{\mathbf{i}} \right] \log(\Mud_{\mathbf{i}\bs{c} k})],
    \end{align}
 where $\Mud_{\mathbf{i}\bs{c}k} := \gamma_{k}^{\ms{(d)}} \sprod_{r=1}^d \theta_{i_r c_r} \mathrm{w}_{c_r k} $.
  At first glance, maximizing $\mathcal{Q}$ is computationally expensive. Evaluating $\mathcal{Q}$ requires computing the conditional expectation of the most fundamental latent subcounts $\Ad_{\mathbf{i}\bs{c}k}$, given by~\looseness=-1
  \begin{align*}
          \Exp\left[\Ad_{\mathbf{i} \bs{c} k} \mid \Ad_{\mathbf{i}}\right] \propto  \gamma_{k}^{\ms{(d)}} \sprod_{r=1}^d \theta_{i_r c_r} \mathrm{w}_{c_r k}, \label{eq:flc}
  \end{align*}
and there are $C^d \cdot K$ of them for each multi-index $(i_1, \dots, i_d)$. This introduces a cost of $\mathrm{O}(K \sum_{d=2}^D ||\Ad||_0 C^d)$ over all hyperedges. However, all that are required are the expected sufficient statistics
\begin{align*}
    \varphi^{\ms{(d)}}_{\mathbf{i}k} &= \sum_{c_1=1}^C \ldots \sum_{c_d=1}^C \Exp\left[\Ad_{\mathbf{i} \bs{c} k} \mid \Ad_{\mathbf{i}}\right],\\ 
    \varphi_{i k}  &= \sum_{d=2}^D \sum_{\mathbf{i} \in \Omegad, i \in \mathbf{i}} \varphi^{\ms{(d)}}_{\mathbf{i}k}, \text{ and }
    \varphi_{ick} = \Exp[\sum_{d=2}^D \sum_{\mathbf{i} \in \Omegad, i \in \mathbf{i}} \sum_{c_r: i_r \neq i}\Ad_{ \mathbf{i} \bs{c} k} \mid \Ad_{\mathbf{i} k}].
\end{align*}

These expectations are available in closed form and cost ${\mathrm{O}(K N_{nz} + N C K)}$ to compute, a cost linear in $C$ and $K$. This cost is typically dominated by the first term, the number of nonzero hyperedges by the number of communities. The M-step iterates through parameters, maximizing $\mathcal{Q}$ with respect to each parameter, conditional on all other parameters fixed. We give closed-form expressions for the optimal $\gamma_{k}^{\ms{(d)}}$ and $\theta_{ic}$ in each of the strictly, semi-, and omniassortative models, as well as derivations of these expressions in the next section. To update $\mathrm{W}$, we use automatic differentiation~\cite{baydin2018automatic}, taking gradient steps to increase $\mathcal{Q}(\Aall, \Muall)$.
 For a user-specified step size $\delta > 0$, we let $\nu_{ck} = \log(\exp(\mathrm{w}_{ck}) - 1) \in \mathbb{R}$ and update $\bs{\nu}$, writing $\mathrm{W} = \mathrm{W}(\bs{\nu})$ as a function of $\bs{\nu}$ and use the update rule~\looseness=-1
\[
    \bs{\nu} \leftarrow \bs{\nu} + \delta \nabla_{\bs{\nu}} \mathcal{Q}(\mathrm{W}(\bs{\nu})),\hspace{1em} \mathrm{w}_{c k} \leftarrow \log(\exp(\nu_{c k})+1).
\]
For sufficiently small $\delta$, the update rule increases $\mathcal{Q}$. 
The cost of evaluating $\nabla_{\bs{v}} \mathcal{Q}$ is equal to the cost of evaluating the subset of $\mathcal{Q}$ dependent on $\bs{\nu}$, given by 
\begin{align*}
    \mathcal{B}(\bs{\nu}) = \Psi(\bs{\nu}) + \sum_{c=1}^C \sum_{k=1}^K \varphi_{ck} \log \left(\mathrm{w}_{c k}(\nu_{c k})\right),
\end{align*}
where 
$\Psi(\bs{\nu}) = - \sum_{d=2}^D \sum_{\mathbf{i} \in \Omegad} \Mud_{\mathbf{i}}(\bs{\nu})$ explicitly denotes the dependence of $\Mud_{\mathbf{i}}$ on $\bs{\nu}$ and ${\varphi_{ck} = \sum_{i=1}^N \varphi_{ick}}$.We derive a dynamic programming algorithm to compute $\Psi(\bs{\nu})$ in $\mathrm{O}(N D K)$ time; each gradient evaluation costs $\mathrm{O}(NDK + KC)$. To preserve the pure community constraint, we keep the first $C$ columns of $\mathrm{W}$ fixed and update the remaining parameters. ~\looseness=-1

\subsection*{Model-specific updates}\label{sec:optimization}
Here, we give specific updates in the E and M steps for each of the strictly assortative, semi-assortative, and omniassortative models. 

\textbf{Strictly assortative.} The model is equivalent to Hypergraph-MT~\cite{contisciani2022inference} parameterized by $C$ communities. In particular, each observed hyperedge may be expressed as the sum of $C$ latent subcounts:
\begin{align*}
    \Ad_{\mathbf{i}} &= \sum_{c=1}^C \Ad_{\mathbf{i}c},\\
    \Ad_{\mathbf{i} c} &\overset{\text{ind.}}\sim \text{Poisson}(\gamma_{c}^{\ms{(d)}} \sprod_{i \in \mathbf{i}} \m_{ic}).
\end{align*}
The evidence lower bound (up to a scalar constant) is 
\begin{align*}
    \mathcal{Q}(\Aall, \Muall) &= - \sum_{d=2}^D \sum_{\mathbf{i} \in \Omegad} \Mud_{\mathbf{i}}  + \sum_{d=2}^D \sum_{\mathbf{i} \in \Omegad}\sum_{c=1}^C \Exp[\Ad_{\mathbf{i}c} \mid \Ad_{\mathbf{i}}]\log \Mud_{\mathbf{i}c} \\
    &= - \sum_{d=2}^D \sum_{\mathbf{i} \in \Omegad}\sum_{c=1}^C \Mud_{\mathbf{i}c}  + \sum_{d=2}^D \sum_{\mathbf{i} \in \Omegad}\sum_{c=1}^C \Exp[\Ad_{\mathbf{i}c} \mid \Ad_{\mathbf{i}}]\log \Mud_{\mathbf{i}c} \\
    &= - \sum_{d=2}^D \sum_{c=1}^C \gamma_{c}^{\ms{(d)}} \sum_{\mathbf{i} \in \Omegad}  \sprod_{i \in \mathbf{i}} \theta_{i c} + \sum_{c=1}^C \sum_{d=2}^D \sum_{\mathbf{i} \in \Omegad} \varphi^{\ms{(d)}}_{\mathbf{i}c}\left(\log \gamma_{c}^{\ms{(d)}} + \sum_{i \in \mathbf{i}} \log \theta_{ic} \right) \\
    &= - \sum_{d=2}^D \sum_{c=1}^C \gamma_{c}^{\ms{(d)}} \phi_{c}^{\ms{(d)}} + \sum_{c=1}^C \sum_{d=2}^D \sum_{\mathbf{i} \in \Omegad} \varphi^{\ms{(d)}}_{\mathbf{i}c} \left(\log \gamma_{c}^{\ms{(d)}} + \sum_{i \in \mathbf{i}} \log \theta_{i c}\right)\\
    &= - \sum_{d=2}^D \sum_{c=1}^C \gamma_{c}^{\ms{(d)}} \phi_{c}^{\ms{(d)}} + \sum_{c=1}^C \sum_{d=2}^D \sum_{\Ad_{\mathbf{i}} > 0} \varphi^{\ms{(d)}}_{\mathbf{i}c} \left(\log \gamma_{c}^{\ms{(d)}} + \sum_{i \in \mathbf{i}} \log \theta_{i c} \right) .
\end{align*}

\textbf{E-step}. Conditional on $\Ad_{\mathbf{i}} > 0$, the vector of latent subcounts $(\Ad_{\mathbf{i} c})_{c=1}^C$ is multinomial distributed, where 
\begin{equation*}
(\Ad_{\mathbf{i} c})_{c=1}^C \mid \Ad_{\mathbf{i}} \sim \text{Multinomial}\left(\Ad_{\mathbf{i}}, \tfrac{\gamma_{c}^{\ms{(d)}} \sprod_{i \in \mathbf{i}} {\theta_{i c}}}{\sum_{c'=1}^C \gamma_{c'}^{\ms{(d)}} \sprod_{i \in \mathbf{i}} \theta_{i c'}}\right).
\end{equation*}
The expected latent subcounts are given by
\be\label{eqn:rhoec}
\varphi^{\ms{(d)}}_{\mathbf{i}c} = \Exp[\Ad_{\mathbf{i}c} \mid \Ad_{\mathbf{i}}] = \Ad_{\mathbf{i}} \cdot \frac{\gamma_{c}^{\ms{(d)}} \sprod_{i \in \mathbf{i}} \theta_{i c}}{\sum_{c'=1}^C \gamma_{c'}^{\ms{(d)}} \sprod_{i \in \mathbf{i}} \theta_{i c'}}.
\ee
We add them together to obtain
$\varphi_{ic} = \sum_{d=2}^D \sum_{\mathbf{i} \in \Omegad} \varphi^{\ms{(d)}}_{\mathbf{i}c}$ and $\varphi_{c}^{\ms{(d)}} = \sum_{\mathbf{i} \in \Omegad} \varphi^{\ms{(d)}}_{\mathbf{i}c}$. 

\textbf{M-step.} The optimal coordinate-wise updates are found by setting the partial derivative (with respect to a given parameter) to $0$ and solving for the given parameter. The optimal updates are given by
\begin{equation}\label{eqn:thetagamma}
    \theta_{ic} = \frac{\varphi_{ic}}{\sum_{d=2}^D \gamma_{c}^{\ms{(d)}} \bar \phi_{ic}^{\ms{(d-1)}}}, \hspace{1em} \gamma_{c}^{\ms{(d)}} = \frac{\varphi_{c}^{\ms{(d)}}}{\phi_{c}^{\ms{(d)}}}, 
\end{equation}
where $\phi_{c}^{\ms{(d)}}$ is defined as $\phi_{c}^{\ms{(d)}} := \sum_{\mathbf{i} \in \Omegad} \sprod_{i \in \mathbf{i}} \theta_{i c}$ and $\bar \phi_{ic}^{\ms{(d)}} := \sum_{\mathbf{i} \in \Omegad, i \not \in \mathbf{i}} \sprod_{j \in \mathbf{i}} \theta_{j c}$. 

\noindent \textbf{Semi-asssortative.} The model is given by 
\begin{align}
    \Ad_{\mathbf{i}} &= \sum_{c_1=1}^C \ldots \sum_{c_d=1}^C \sum_{k=1}^K \Ad_{\mathbf{i} \mathbf{c} k},\\
    \Ad_{\mathbf{i} \mathbf{c} k} &\overset{\text{ind.}}{\sim} \text{Poisson}(\gamma_{k}^{\ms{(d)}} \sprod_{q=1}^d (\mathrm{w}_{c_q k}\theta_{i_q c_q})). 
\end{align}
The EM algorithm is given in \Cref{alg:em}~and alternates between the E-step, computing the expectations conditional on $(\mathrm{\Gamma}, \mathrm{W}, \mathrm{\Theta})$ using \Cref{eq:e1} and \Cref{eq:e2}, and the M-step, maximizing $\mathcal{Q}(\Aall, \Muall)$ with respect to the parameters $(\mathrm{\Gamma}, \mathrm{W}, \mathrm{\Theta})$. 
 
For $\mathbf{i} = (i_1, \dots, i_d)$, the conditional expectation ${\Exp[\Ad_{\mathbf{i} k}\mid \Ad_{\mathbf{i}}] = \varphi^{\ms{(d)}}_{\mathbf{i}k}}$ is given by 
\begin{align}
 \varphi^{\ms{(d)}}_{\mathbf{i} k} = \Ad_{\mathbf{i}} \cdot \frac{\gamma_{k}^{\ms{(d)}} \sprod_{i \in \mathbf{i}} \bs{\theta}_{i}^\top\mathbf{w}_k}{\sum_{k'=1}^K \gamma_{k'}^{\ms{(d)}} \sprod_{i \in \mathbf{i}} \bs{\theta}_{i}^\top\mathbf{w}_{k'}}\label{eq:e1} ,
\end{align}
under the multinomial thinning property of the Poisson.
Then, we have:
\begin{align}
  \Exp[\Ad_{i c k}\mid \Aall] =   \varphi^{\ms{(d)}}_{i c k} = \frac{\theta_{i c} \mathrm{w}_{c k}}{\sum_{c'=1}^C \theta_{i c'}\mathrm{w}_{ c' k}} \cdot \sum_{d=2}^D \sum_{\mathbf{i} \in \Omegad} \varphi^{\ms{(d)}}_{\mathbf{i}k} 1(i \in \mathbf{i}). \label{eq:e2}
\end{align}

\begin{algorithm}[H] 
\caption{EM algorithm for the semi-assortative model.} 
\begin{algorithmic}\label{alg:em}
    \Require Adjacency tensors $\Aall$, initialized parameters $\mathrm{\Gamma}, \mathrm{W}, \mathrm{\Theta}$, step size $\delta > 0$, stopping criterion. 
    \Repeat 
        \State \textbf{E-step}: compute the expected latent subcounts as in \Cref{eq:e1,eq:e2}
        \[
        \varphi^{\ms{(d)}}_{\mathbf{i}k} \leftarrow \Ad_{\mathbf{i}} \cdot \frac{\gamma_{k}^{\ms{(d)}} \sprod_{i \in \mathbf{i}}  \bs{\theta}_{i}^\top\mathbf{w}_k}{\sum_{k'=1}^K \gamma_{k'}^{\ms{(d)}} \sprod_{i \in \mathbf{i}} \bs{\theta}_{i}^\top\mathbf{w}_{k'}}
        \]
        \[
        \varphi_{ik} \leftarrow \sum_{d=2}^D \sum_{\mathbf{i} \in \Omegad, i \in \mathbf{i}} \varphi^{\ms{(d)}}_{\mathbf{i}k}, \hspace{2em} \varphi_{ick} \leftarrow \varphi_{ik} \cdot \frac{\mathrm{w}_{ck}\theta_{ic}}{\sum_{c'=1}^C \mathrm{w}_{c' k} \theta_{ic'}}
        \]

        \State \textbf{M-step}: update the parameters as in \Cref{eqn:thetagamma}
        \[
        \gamma_{k}^{\ms{(d)}} \leftarrow \frac{\sum_{\mathbf{i} \in \Omegad} \varphi^{\ms{(d)}}_{\mathbf{i}k}}{\phi_{k}^{\ms{(d)}}}
        \]
        \[\textbf{for } i \in [N]:
        \theta_{ic} \leftarrow \frac{\sum_{k=1}^K \varphi_{ick}}{\sum_{k=1}^K \mathrm{w}_{c k} \left(\sum_{d=2}^D \gamma_{k}^{\ms{(d)}} \bar \phi_{ik}^{\ms{(d-1)}}\right)}
        \]
        \hspace{19 em} update $\bs{\phi}, (\bar \phi_{i+1, k}^{\ms{(d)}})_{d=2}^D$
        \[
        \bs{\nu} \leftarrow \bs{\nu} + \delta \nabla \mathcal{Q}(\mathrm{W}(\bs{\nu})), \hspace{1em} \mathrm{W} \leftarrow \log(\exp(\bs{\nu}) + 1)
        \]
        \hspace{1.5em} compute $\bs{\phi}, (\bs{ \bar \phi}^{\ms{(d)}})_{d=2}^D$
    \Until{stopping criterion met}
    \State \Return $\mathrm{\Gamma}, \mathrm{W}, \mathrm{\Theta}$
\end{algorithmic}
\end{algorithm}

\textbf{Omniassortative.}
In Supplementary Note 4, we show that the semi-assortative model is constrained to satisfy the inequality 
\begin{align}\label{lemma:prop}
    \frac{\sum_{c=1}^C \sum_{k=1}^K \Lambdad_{c \dots c k}}{\sum_{c_1=1}^C  \ldots \sum_{c_d=1}^C \sum_{k=1}^K \Lambdad_{c_1 \dots c_d k}} \geq \frac{1}{C^{d-1}} ,
\end{align}
where $\Lambdad_{c_1 \dots c_d k} = \gamma_{k}^{\ms{(d)}} \sprod_{q=1}^d \mathrm{w}_{c_q k}$, such that $\Lambdad_{c_1 \dots c_d} = \sum_{k=1}^K \Lambdad_{c_1 \dots c_d k}$. To relax the bound, we impose that for $k > C$, the diagonal elements $\Lambdad_{c \dots c k} = 0$. Under this modification, we express each affinity tensor as
\begin{equation} \label{eq:omni-core}
\Lambdad_{c_1 \dots c_d k} = 1(k \leq C \text{ or }\exists c_i \neq c_j)\gamma_{k}^{\ms{(d)}} \sprod_{q=1}^d \mathrm{w}_{c_q k}.
\end{equation}
Under this modification, inequality~(\ref{lemma:prop}) is not necessarily satisfied and we may capture pure disassortativity (i.e., $\mathrm{\Lambda}_{c_1 c_1}^{\ms{(2)}} \approx 0$ but $\mathrm{\Lambda}_{c_1 c_2}^{\ms{(2)}} > > 0$). Under this adaptation, we derive a similar generalized EM algorithm with closed-form updates to $\mathrm{\Theta}$ and $\mathrm{\Gamma}$ as in the semi-assortative setting. The adaptation requires an additional factor of $C$ in the computational cost of the E-step; see~\Cref{alg:emO} for more details. 

The evidence lower bound is
 \begin{align*}
     \mathcal{Q}(\Aall, \Muall) =  \sum_{d=2}^D \sum_{\mathbf{i} \in \Omegad} \sum_{k=1}^K \sum_{c_1=1}^C \ldots \sum_{c_d=1}^C  \left(1(k \leq C \text{ or }\exists c_i \neq c_j)\gamma^{\ms{(d)}}_k \sprod_{r=1}^d (\theta_{i_r c_r} \mathrm{w}_{c_r k} ) + \Ad_{\mathbf{i} \bs{c} k } \left(\log \gamma_{k}^{\ms{(d)}} + \sum_{r=1}^d\log(\theta_{i_r c_r}\mathrm{w}_{c_r k})\right)\right).
 \end{align*}
  The optimal update for $\theta_{ic}$ is
 \begin{align}
     \theta_{ic} = \frac{\sum_{k=1}^K \varphi_{ick}}{\sum_{d=2}^D \gamma_{c}^{\ms{(d)}} \bar \phi_{ic}^{\ms{(d-1)}} + \sum_{k=C + 1}^K \gamma_{k}^{\ms{(d)}} \mathrm{w}_{ck}\left(\bar \phi_{ik}^{\ms{(d-1)}} - \mathrm{w}_{ck}^{d-1} \bar \phi_{ic}^{\ms{(d-1)}}\right)}  . \label{eqn:thetadis}
 \end{align}
The optimal update for $\gamma_{c}^{\ms{(d)}}, \, c \in [C]$ is 
 \begin{align}
     \gamma_{c}^{\ms{(d)}} = \frac{\sum_{\mathbf{i} \in \Omegad} \varphi^{\ms{(d)}}_{\mathbf{i}c}}{\phi_{c}^{\ms{(d)}}} ,\label{eqn:gammacdis}
 \end{align}
 and the optimal update for $\gamma_{k}^{\ms{(d)}}, \, k > C$ is 
 \begin{align}
     \gamma_{k}^{\ms{(d)}} = \frac{\sum_{\mathbf{i} \in \Omegad} \varphi^{\ms{(d)}}_{\mathbf{i}k}}{\phi_{k}^{\ms{(d)}} - \sum_{c=1}^C \mathrm{w}_{ck}^d \phi_{c}^{\ms{(d)}}}.\label{eqn:gammakdis}
 \end{align}
 Moreover, 
 \begin{align}
      \varphi_{ick} = \sum_{d=2}^D \sum_{\mathbf{i} \in \Omegad} \varphi^{\ms{(d)}}_{\mathbf{i}ick},\label{eqn:rhoick}
 \end{align}
 where 
 \begin{align}
\varphi^{\ms{(d)}}_{\mathbf{i}ick} &= \varphi_{\mathbf{i}k}^{\ms{(d)}} \cdot \frac{ \theta_{ic} \mathrm{w}_{ck}\left(\sprod_{j \neq i} m_{j k} - \mathrm{w}_{ck}^{d-1} \sprod_{j \neq i} \theta_{j c}\right)}{\sum_{c'=1}^C \mathrm{w}_{c'k} \theta_{ic'}\left(\sprod_{j \neq i} m_{j k} - \mathrm{w}_{c'k}^{d-1} \sprod_{j \neq i} \theta_{j c'}\right)},\label{eqn:rhoicke} \\
\varphi^{\ms{(d)}}_{\mathbf{i} k} &= \Ad_{\mathbf{i}} \cdot \frac{\sprod_{i \in \mathbf{i}}\mathbf{w}^\top_{k'} \bs{\theta}_i - 1(k > C)\sum_{c=1}^C \mathrm{w}_{ck}^d \sprod_{i \in \mathbf{i}} \theta_{ic}}{ \sum_{k'=1}^K \left( \sprod_{i \in \mathbf{i}}\mathbf{w}^\top_{k'} \bs{\theta}_i - 1(k' > C)\sum_{c=1}^C \mathrm{w}_{ck'}^d \sprod_{i \in \mathbf{i}} \theta_{ic}\right)}. \label{eqn:rhoike}
 \end{align}

 \begin{algorithm}[H] 
\caption{EM algorithm for the omniassortative model.} 
\begin{algorithmic}\label{alg:emO}
    \Require Adjacency tensors $\Aall$, initialized parameters $\mathrm{\Gamma}, \mathrm{W}, \mathrm{\Theta}$, step size $\delta > 0$, stopping criterion. 
    \Repeat 
        \State \textbf{E-step}: $M \leftarrow \mathrm{\Theta} \mathrm{W}$, compute the expected latent subcounts as in \Cref{eqn:rhoick,eqn:rhoike,eqn:rhoicke}
        \[
\varphi^{\ms{(d)}}_{\mathbf{i}k} \leftarrow \Ad_{\mathbf{i}} \cdot \frac{\sprod_{i \in \mathbf{i}}\mathbf{w}^\top_{k'} \bs{\theta}_i - 1(k > C)\sum_{c=1}^C \mathrm{w}_{ck}^d \sprod_{i \in \mathbf{i}} \theta_{ic}}{ \sum_{k'=1}^K \left( \sprod_{i \in \mathbf{i}}\mathbf{w}^\top_{k'} \bs{\theta}_i - 1(k' > C)\sum_{c=1}^C \mathrm{w}_{ck'}^d \sprod_{i \in \mathbf{i}} \theta_{ic}\right)},
        \]
        \[\varphi^{\ms{(d)}}_{\mathbf{i}ick} \leftarrow \frac{ \theta_{ic} \mathrm{w}_{ck}\left(\sprod_{j \neq i} m_{j k} - \mathrm{w}_{ck}^{d-1} \sprod_{j \neq i} \theta_{j c}\right)}{\sum_{c'=1}^C \mathrm{w}_{c'k} \theta_{ic'}\left(\sprod_{j \neq i} m_{j k} - \mathrm{w}_{c'k}^{d-1} \sprod_{j \neq i} \theta_{j c'}\right)} \cdot \varphi^{\ms{(d)}}_{\mathbf{i}k}, \quad
        \varphi_{ick} \leftarrow \sum_{d=2}^D \sum_{\mathbf{i} \in \Omegad} \varphi^{\ms{(d)}}_{\mathbf{i}ick}
        \]

        \State \textbf{M-step}: update the parameters as in \Cref{eqn:thetadis,eqn:gammacdis,eqn:gammakdis}
        \[
        \gamma_{k}^{\ms{(d)}} \leftarrow \frac{\sum_{\mathbf{i} \in \Omegad} \varphi^{\ms{(d)}}_{\mathbf{i}k}}{\phi_{k}^{\ms{(d)}} - \sum_{c=1}^C \mathrm{w}_{ck}^d \phi_{c}^{\ms{(d)}}}, \quad \gamma_{c}^{\ms{(d)}} \leftarrow  \frac{\sum_{\mathbf{i} \in \Omegad} \varphi^{\ms{(d)}}_{\mathbf{i}c}}{\phi_{c}^{\ms{(d)}}}
        \]
        \[\textbf{for } i \in [N]:
        \theta_{ic} \leftarrow \frac{\sum_{k=1}^K \varphi_{ick}}{\sum_{d=2}^D \gamma_{c}^{\ms{(d)}} \bar \phi_{ic}^{\ms{(d-1)}} + \sum_{k=C + 1}^K \gamma_{k}^{\ms{(d)}} \mathrm{w}_{ck}\left(\bar \phi_{ik}^{\ms{(d-1)}} - \mathrm{w}_{ck}^{d-1} \bar \phi_{ic}^{\ms{(d-1)}}\right)}
        \]
        \hspace{19 em} update $\bs{\phi}, (\bar \phi_{i+1, k}^{\ms{(d)}})_{d=2}^D$
        \[
        \bs{\nu} \leftarrow \bs{\nu} + \delta \nabla \mathcal{Q}(\mathrm{W}(\bs{\nu})), \hspace{1em} \mathrm{W} \leftarrow \log(\exp(\bs{\nu}) + 1)
        \]
        \hspace{1.5em} compute $\bs{\phi}, (\bs{ \bar \phi}^{\ms{(d)}})_{d=2}^D$
    \Until{stopping criterion met}
    \State \Return $\mathrm{\Gamma}, \mathrm{W}, \mathrm{\Theta}$
\end{algorithmic}
\end{algorithm}

\section*{Supplementary Note 2: Additional mathematical details}

\subsection*{Priors and MAP estimation}
We may consider maximum a posteriori (MAP) estimation as an alternative to maximum likelihood estimation to infer the parameters.
Here we provide a framework for this, by placing Gamma priors on the elements $\theta_{ic}$ of the node class-membership matrix and the community rate terms $\gamma_{k}^{\ms{(d)}}$. The probability density function of $\gamma \sim \text{Gamma}(\alpha, \beta) $ is given by 
\begin{align}
    \text{Gamma}(\gamma; \alpha, \beta) = \frac{\beta^\alpha}{\mathrm{\Gamma}(\alpha)} \gamma^{\alpha-1} e^{- \beta \gamma} .
\end{align}
In the case of independent and identically distributed $\theta_{ic} \overset{i.i.d.}{\sim} \text{Gamma}(\alpha, \beta)$,
the log likelihood is extended from $\mathcal{L}$ to $\mathcal{L} - (\alpha - 1)\sum_{i, c} \log(\theta_{ic}) - \beta \sum_{i, c} \theta_{ic}$. 

Under the maximum likelihood formulation, $\mathrm{y} \sim \text{Poisson}(c \gamma)$, the optimal update is $\gamma = \frac{\mathrm{y}}{c}$ for scalar $c \in \mathbb{R}_{>0}$. Under the gamma prior $\gamma \sim \text{Gamma}(\alpha, \beta)$, the optimal update is ${\gamma = \max(\frac{\mathrm{y} + \alpha - 1}{c + \beta}, 0)}$. That is, we add $\alpha - 1$ to the numerator and $\beta$ to the denominator, and ensure the resulting update is non-negative. When $\alpha\!=\!1$, the gamma distribution simplifies to the Exponential$(\beta)$ distribution, and the optimal update is $\gamma\!= \!\frac{\mathrm{y}}{c + \beta}$. We can apply these rules to the closed-form updates under each model. We note that such a parameterization removes the constraint that $||\bs{\theta}_{c}||=1$.~\looseness=-1

\subsection*{Graph (pairwise) setting}
In the standard (pairwise) graph setting, all hyperedges are pairwise, i.e., $d\!=\!2$. Then
\begin{align}
    \Mud_{ij} = \sum_{k=1}^K \gamma_{k}^{\ms{(2)}}  \sum_{c_1 =1}^C \sum_{c_2=1}^C \mathrm{w}_{c_1 k} \mathrm{w}_{c_2 k} \theta_{i c_1} \theta_{i c_2} = \sum_{c_1 =1}^C \sum_{c_2=1}^C \tilde{\mathrm{w}}_{c_1 c_2} \theta_{i c_1} \theta_{i c_2} \,  ,\label{eqn:tucker2}
    \end{align}
 where $\tilde{\mathrm{w}}_{c_1 c_2}:=\sum_{k=1}^K \gamma_{k}^{\ms{(2)}}  \mathrm{w}_{c_1 k} \mathrm{w}_{c_2 k}$. We thus obtain a Tucker-2 decomposition \cite{kolda_tensor_2009}, which has been used in various mixed-membership models for networks~\cite{ball2011efficient,de_bacco_community_2017,schein_bayesian_2016,contisciani2020community} to model disassortative community structure.
 
We can also find an alternative factorization, by writing the matrix $M = \mathrm{\Theta} \mathrm{W}$ with entries $m_{ik}=\sum_{c=1}^C \,\theta_{ic} \mathrm{w}_{ck}$. With this, we get 
\begin{align*}
    \Mud_{ij} &= \sum_{k=1}^K \gamma_{k}^{\ms{(2)}}   \sum_{c_1=1}^C (\theta_{ic_1} \mathrm{w}_{c_1 k}) \sum_{c_2=1}^C ( \theta_{jc_2} \mathrm{w}_{c_2 k})\\
    &= \sum_{k=1}^K\gamma_{k}^{\ms{(2)}}  \, (\bs{\theta}_{i}^\top \mathbf{w}_{k}) (\bs{\theta}_{j}^\top \mathbf{w}_{k} )
    = \sum_{k=1}^K \gamma_{k}^{\ms{(2)}}   m_{ik} m_{jk}.
\end{align*}
This is a symmetric non-negative matrix factorization that generalizes into a canonical polyadic (CP) decomposition~\cite{kolda_tensor_2009, hitchcock1927expression} if we consider $\bs{\gamma}^{\ms{(2)}}$ to be the diagonal entries of a diagonal $K \times K$ affinity matrix. It is an assortative decomposition over the classes $[C]$, but allows for disassortativity over the nodes. In fact, writing $S = \text{diag}(\bs{\gamma}^{\ms{(2)}})$, $\Muall = M S M^\top = \mathrm{\Theta} \mathrm{W} S \mathrm{W}^\top \mathrm{\Theta}^\top \in \mathbb{R}^{N \times N}$, we obtain a bilinear model with affinity matrix $\tilde{\mathrm{W}}: =\mathrm{W} S \mathrm{W}^\top$ as in \Cref{eqn:tucker2}.

\subsection*{Properties of the Poisson}
In our derivations we used two main properties of the Poisson distribution~\cite{kingman1992poisson}, as described below.

\textbf{Poisson additivity.}\label{lem:poisson-add}
 Let $y_k \overset{\text{ind.}}{\sim} \text{Poisson}(\lambda_k)$ for $k \in [K]$. Then marginally, the sum ${y = \sum_{k=1}^K y_k \sim \text{Poisson}(\lambda)}$ is also Poisson distributed, with rate $\lambda = \sum_{k=1}^K \lambda_k$.~\looseness=-1

\textbf{Multinomial thinning.}\label{lem:pm} 
    Let $y_k \overset{\text{ind.}}{\sim} \text{Poisson}(\lambda_k)$ for $k \in [K]$. Then conditional on the sum $y_\bullet = \sum_{k=1}^K y_k = n$, the vector $\bs{y} = (y_k)_{k=1}^K$ is distributed as 
    \begin{align*}
            (\bs{y} \mid y_{\bullet} = n) \sim \begin{cases}
            \text{Multinomial}\left(n, \bs{\pi}\right), \text{ where } \pi_k = \frac{\lambda_k}{\sum_{k'=1}^K \lambda_{k'}} &\text{ if } n > 0\\
            \delta_{\bs{0}} &\text{otherwise}.
            \end{cases}
    \end{align*}

\subsection*{Computation}
The updates to $\gamma^{\ms{(d)}}_k$, $\theta_{i c}$ and $\mathrm{w}_{c k}$ require computing the sums over a combinatorial number of summands
\begin{align*}
    \phi_{k}^{\ms{(d)}} := \sum_{\mathbf{i}: |\mathbf{i}| = d} \sprod_{r=1}^d \bs{\theta}_{i_r}^\top \mathbf{w}_k = \sum_{\mathbf{i}: |\mathbf{i}| = d} \sprod_{r=1}^d m_{i_r k} ,
\end{align*} 
and 
\begin{align*}
\bar \phi^{\ms{(d)}}_{i k} := \sum_{\mathbf{i}: |\mathbf{i}| = d, i \not \in \mathbf{i}} \sprod_{r=1}^d \bs{\theta}_{i_r}^\top \mathbf{w}_k = \sum_{\mathbf{i}: |\mathbf{i}| = d, i \not \in \mathbf{i}} \sprod_{r=1}^d m_{i_r k}. 
\end{align*}

In particular, each $\phi_{k}^{\ms{(d)}}$ contains ${N \choose d}$ summands and each $\bar \phi_{i k}^{\ms{(d)}}$ contains ${N - 1 \choose d}$ summands. Since the optimal $\theta_{i c}$ depends on the values of $\theta_{i' c'}$ for other nodes, updates to $\theta_{i c}$ are performed sequentially. Each update to $\theta_{i c}$ changes $\phi^{\ms{(d)}}_{i' k}$ for $i' \neq i$, necessitating a recalculation of $\bar \phi^{\ms{(d)}}_{ik}$ for each $i$ in a sequential manner. We use the algorithm derived in \cite{contisciani2022inference} to compute each term in $\mathrm{O}(1)$ time, which conditional on the previous values, updates $\phi_{k}^{\ms{(d)}}$ according to an update rule.~\looseness=-1

The updates to $\mathrm{W}$, however, cannot rely on this algorithm to compute $\phi_{k}^{\ms{(d)}}$, as the automatic differentiation call requires a function which computes $\phi_{k}^{\ms{(d)}}$ as functions of $\mathrm{\Theta}$ and $\mathrm{W}$ from scratch. Therefore, we derive a dynamic programming algorithm to compute $\phi_{k}^{\ms{(d)}}$ and $\bar \phi^{\ms{(d)}}_{ik}$ efficiently. Letting $\bar \phi^{\ms{(0)}}_{i k} = 1$, it holds that
\begin{align}\label{def:phi}
    \phi_{k}^{\ms{(d)}} = \frac{1}{d}\sum_{i=1}^N m_{i k} \bar \phi^{\ms{(d-1)}}_{i k},\hspace{1em} 
    \bar \phi^{\ms{(d)}}_{ik} = \phi_{k}^{\ms{(d)}} - m_{i k} \bar \phi_{i k}^{\ms{(d-1)}}\quad\text{for $d \in [D]$.}
\end{align}

\begin{proof}
    The result relies on the following proposition. 
    
    \noindent \textbf{Proposition.}
        As defined in~\Cref{def:phi}, $\bar \phi_{ik}^{\ms{(d-1)}}$ may be equivalently expressed as $\bar \phi_{ik}^{\ms{(d-1)}} = \sum_{\mathbf{i} \in \Omegad, i \in \mathbf{i}} \sprod_{j \neq i} m_{jk}$.
    \begin{proof}
Let  
\[
\Omegad_i = \{\mathbf{i} \in \Omegad: i \in \mathbf{i} \},  
\quad  
\bar{\Omega}^{\ms{(d)}}_i = \{ \mathbf{i} \in \Omegad: i \not \in \mathbf{i} \}.
\]  
Define the bijection  
\[
f_i: \bar{\Omega}^{\ms{(d-1)}}_i \to \Omegad_i, \quad f_i(\mathbf{i}) = \mathbf{i} \cup \{i\}.
\]  
Since $ f_i $ is bijective, we have  
\[
|\bar{\Omega}^{\ms{(d-1)}}_i| = |\Omegad_i|.
\]  
Moreover, for any $ \mathbf{i} \in \bar{\Omega}^{\ms{(d-1)}}_i $:  
\[
\sprod_{j \in \mathbf{i}} m_{jk} = \sprod_{j \in f_i(\mathbf{i}), j \neq i} m_{jk},
\]  
since $ f_i(\mathbf{i}) \setminus \{i\} =\mathbf{i} \cup \{i\} \setminus \{i\} = \mathbf{i} $.

Therefore,  
\begin{align*}
    \bar \phi_{ik}^{\ms{(d-1)}} &:= \sum_{\mathbf{i} \in \Omega^{\ms{(d-1)}}, i \not \in \mathbf{i}} \sprod_{j \in \mathbf{i}} m_{jk}
    = \sum_{\mathbf{i} \in \bar{\Omega^{\ms{(d-1)}}}_i} \sprod_{j \in \mathbf{i}} m_{jk}  
= \sum_{\mathbf{i} \in \bar{\Omega^{\ms{(d-1)}}}_i} \sprod_{j \in f_i(\mathbf{i}), j \neq i} m_{jk} \\
    &= \sum_{f_i(\mathbf{i}) \in \Omegad_i} \sprod_{j \in f_i(\mathbf{i}), j \neq i} m_{jk} 
    = \sum_{\bs{u} \in \Omegad_i} \sprod_{j \in \bs{u}, j \neq i} m_{jk}  \\
    &= \sum_{\bs{u} \in \Omegad, i \in \bs{u}} \sprod_{j \neq i} m_{jk}.
\end{align*}
    \end{proof}
    \textit{Proof of first relation.} By definition, 
    \begin{align}
    \phi_{k}^{\ms{(d)}} &= \sum_{\mathbf{i} \in \Omegad} \sprod_{i_r=1}^d m_{i_rk}\\
    &= \sum_{\mathbf{i} \in \Omegad} \frac{1}{d} \sum_{r=1}^d \sprod_{i_{r'}=1}^d m_{i_{r'}k}\\
    &= \frac{1}{d} \sum_{\mathbf{i} \in \Omegad}\sum_{r=1}^d m_{i_r k}\sprod_{r' \neq r} m_{i_{r'}k}. \label{eq:phi_dk_def}
    \end{align}
    Since each node $i$ occurs in each hyperedge $\mathbf{i}$ at most once,
    \begin{align}
        \sum_{r=1}^d m_{i_r k}\sprod_{m' \neq m} m_{i_r'k} &= \sum_{r=1}^d \sum_{i=1}^N 1(i_r = i)m_{ik} \sprod_{j \in \mathbf{i}, j \neq i} m_{jk}\\
        &=\sum_{i=1}^N 1(i \in \mathbf{i})m_{ik} \sprod_{j \in \mathbf{i}, j \neq i} m_{jk}.\label{eq:phi_dk_new}
    \end{align}
    Plugging \Cref{eq:phi_dk_new} into \Cref{eq:phi_dk_def} yields 
    \begin{align*}
    \phi_{k}^{\ms{(d)}} &= \frac{1}{d}\sum_{i=1}^N m_{ik} \sum_{\mathbf{i} \in \Omegad, i \in \mathbf{i}} \sprod_{j \in \mathbf{i}, j \neq i} m_{jk}\\
    &= \frac{1}{d}\sum_{i=1}^N m_{ik} \sum_{\mathbf{i} \in \Omega^{\ms{(d-1)}}, i \not \in \mathbf{i}} \sprod_{j \in \mathbf{i}} m_{jk}\\
    &= \frac{1}{d}\sum_{i=1}^N m_{ik} \bar \phi_{ik}^{\ms{(d-1)}}. 
    \end{align*}
    \textit{Proof of second relation.}
    For $i \in [N]$, we partition $\Omegad$ into two sets, based on whether a hyperedge contains node $i$.  
    \be
    \Omegad = \{\mathbf{i} \in \Omegad: i \in \mathbf{i}\} \cup \{\mathbf{i} \in \Omegad: i \not\in \mathbf{i}\} .
    \ee 
    Then the sum over all hyperedges of order $d$ expands as 
    \begin{align*}
    \phi_{k}^{\ms{(d)}} &= \sum_{\mathbf{i} \in \Omegad} \sprod_{r=1}^d m_{i_rk}
    = \sum_{\mathbf{i} \in \Omegad: i \in \mathbf{i}} \sprod_{r=1}^d m_{i_rk} + \sum_{\mathbf{i} \in \Omegad: i \not \in \mathbf{i}} \sprod_{r=1}^d m_{i_rk}\\
    &= m_{ik} \phi^{\ms{(d-1)}}_{ik} + \bar \phi^{\ms{(d)}}_{ik},
    \end{align*}
\end{proof}
implying that $\bar \phi_{ik}^{\ms{(d)}} = \phi_{k}^{\ms{(d)}} - m_{ik} \bar \phi_{ik}^{\ms{(d-1)}}$.
Computation scales as $\mathrm{O}(NDK)$, where $\phi_{i k}^{\ms{(d)}}$ can be computed in parallel over $i,k$ and $\phi_{k}^{\ms{(d)}}$ in parallel over $k$.

\section*{Supplementary Note 3: Experimental details}
\textbf{Initialization.} In our experiments, we initialize the main parameters as:
\begin{align*}
\bs{\theta}_{i} &\overset{\text{i.i.d.}}{\sim} \text{Dirichlet}(10^3,\dots,10^3) \nonumber & \forall i \in [N]\\
\mathbf{w}_k &\overset{\text{i.i.d.}}{\sim} \text{Dirichlet}(1,\dots,1)  &\text{for $k > C$}\\
\gamma_{k}^{\ms{(d)}}&= 1 &\forall d \in \{2, \dots, D\} ,k \in [K] \nonumber.
\end{align*}
\\
To promote the learning of disassortative structure in the experiments in the \texttt{hospital} interaction setting, we initialize $\gamma_{k}^{\ms{(d)}} = 0.01$ for $k \leq C$ and $\gamma_{k}^{\ms{(d)}} = 1$ otherwise. 

\textbf{Training.} All of our experiments were run using one CPU.
Convergence was assessed by monitoring the log-likelihood given in \Cref{eq:llk}.
We trained each model until its value changed by less than 1 over $10$ iterations, or 1,000 iterations, whichever occurred first. The models which require a learning rate were trained with a learning rate of $10^{-6}$. We do not update the first $C$ columns of $\mathrm{W}$ during gradient ascent to preserve the structure $\mathrm{W} = [\mathrm{I}_C \mid \mathbf{w}_{c+1}, \dots ,\mathbf{w}_{K} ]$. ~\looseness=-1

While larger choices of $K$ add computational cost to fitting each model, we find that it is better to overestimate $K$ than underestimate when evaluating results. In particular, we find that if a community contains no meaningful structure, the model learns to set $\gamma^{\mathsmaller{(d)}}_k$ very close to zero for each $d$. ~\looseness=-1

\textbf{Post-training transformation.}
 Each generalized EM algorithm converges to yield estimates $ \mathrm{\Gamma}, \mathrm{\Theta}, \mathrm{W}$. These estimates are not ensured to fall into the constrained model class. Therefore, define the constants and parameters 
 $\psi_c = \sum_{i=1}^N \theta_{ic}, \tilde \theta_{ic} = \frac{\theta_{ic}}{\psi_{c}}, \psi_k = \sum_{c=1}^C \mathrm{w}_{ck} \psi_c, \, \tilde {\mathrm{w}}_{ck} = \frac{\mathrm{w}_{ck} \psi_{c}}{\psi_k}$, $\tilde \gamma_{k}^{\ms{(d)}} = \gamma_{k}^{\ms{(d)}} (\psi_k)^d$. 
 Then $(\tilde {\mathrm{\Gamma}}, \tilde {\mathrm{\Theta}}, \tilde {\mathrm{W}})$ belongs in the constrained model class. Moreover,
\begin{align*}
    \tilde \Capmu^{\ms{(d)}}_{i_1 \dots i_d} &= \sum_{k=1}^K \tilde \gamma_{k}^{\ms{(d)}} \sprod_{r=1}^d \left(\sum_{c=1}^C \tilde {\mathrm{w}}_{ck} \tilde \theta_{ic}\right)\\
    &= \sum_{k=1}^K \gamma_{k}^{\ms{(d)}} (\psi_k)^d
    \sprod_{r=1}^d \left(\sum_{c=1}^C  \frac{\mathrm{w}_{ck} \psi_c \theta_{ic}}{\psi_c \psi_k}\right)\\
    &= \sum_{k=1}^K \gamma_{k}^{\ms{(d)}} \sprod_{r=1}^d \left(\sum_{c=1}^C \mathrm{w}_{ck} \theta_{ic} \right) = \Mud_{i_1 \dots i_d}.
\end{align*}

That is, we may fit the unconstrained model and transform the parameters accordingly only once at the end, after convergence is achieved. Doing so makes them more interpretable and ensures they lie in the identifiable model class. In practice, however, we find that applying these transformations at each iteration improves the numerical stability of the algorithm.

\textbf{Drug case study: selection criteria.}

For $C, K \in (8, 16, 32, 48, 64)$ and $C \leq K$, we fit the omniassortative model to the data, masking a subset of the data and select the $C$ and $K$ that achieves the highest heldout log-likelihood, which occurs for $C\!= \!16$ and $K\!=\!48$ 
(approximately the best combination of $C$ and $K$).~\looseness=-1

To measure how much a community differs from the set of classes, we compute the Jensen-Shannon divergence~\cite{lin1991divergence} from its nearest class, where larger values imply greater values of disassortativity. During inference, our algorithm allocates observed hyperedges to communities and clusters. We can understand the number of hyperedges allocated to a given community as a measure of its importance~\cite{schein2019allocative, yildirim2020bayesian}. Communities with larger assigned latent counts account for more of the observed data.~\looseness=-1

We select two communities with large dissimilarity measures (as measured by Jensen-Shannon divergence to its nearest class) that are assigned at least $1,000$ hyperedge counts (see Methods: allocation). These communities correspond to $k\!=\!19$ and $k\!=\!44$, shown in the main paper. We use these two communities and their corresponding classes to identify related classes and communities (classes are shown in maroon, communities in bronze). The term $\mathrm{w}_{ck}$ for class $c$ and community $k$ is proportional to the width of the edge between class $c$ and community $k$. ~\looseness=-1

 \textbf{Hypergraph generation: inclusion occurrences.}
 The inclusion occurrence of a hyperedge size $d$ is the number of nonzero hyperedges of size $d$ which appear as a subset of a hyperedge of size $d+1$. Computing the inclusion occurrences of the hypergraph we generate is expensive due to the size of the hypergraph. Therefore, for each hyperedge order, we randomly sample $10^4$ hyperedges to form a smaller hypergraph and compute its inclusion occurrences. We repeat this procedure 10 times, for both the true and synthetic hypergraph data and show the average.

\section*{Supplementary Note 4: Theoretical properties}\label{sec:theory}
Here we derive in more details the main theoretical properties described in the main manuscript. We first give a summary of the main results in \Cref{sec:summary} and then provide detailed proofs of each result in \Cref{sec:proofs}. Most of these results focus on the semi-assortative model. \Cref{lem:dis-unique} lets us apply these results to the omniassortative model. Throughout, we assume the $C \leq K \leq N$ and the matrix $\mathrm{\Theta} \in \mathbb{R}^{N \times C}$ is of (full) rank $C$, which holds almost surely. Our measure-theoretic results are with respect to the Lebesgue measure over the parameter space $\mathbb{R}_{> 0}^{D - 1 \times K} \times (\mathrm{\Delta}^{C-1})^{K-C} \times (\mathrm{\Delta}^{N-1})^C$ which contains the parameters $(\gamma^{\ms{(d)}}_k)_{d,k}$, $\mathbf{w}_{c+1}, \dots, \mathbf{w}_{K}$ and $\mathrm{\Theta}$. For the omniassortative model, we consider negative values of $\gamma^{\ms{(d)}}_c, c \in [C]$, as specified below.  

\renewcommand\thesubsection{A}
\setcounter{subsection}{1} 
\renewcommand{\thesection}{A}
\setcounter{section}{0} 

\subsection{Summary of main theoretical results}\label{sec:summary}
Recall that the expression for $\Mud_{i_1 \dots i_d}$ is given by 
\begin{align*}
  \Mud_{i_1 \dots i_d} = \sum_{c_1=1}^C \ldots \sum_{c_d=1}^C  \Lambdad_{c_1 \dots c_d} \sprod_{r=1}^d \theta_{i_r c_r} . 
\end{align*}
The parameterization of $\Lambdad$ is defined element-wise by 
\begin{align*}
\Lambdad_{c_1 \dots c_d} = \sum_{k=1}^{K} \gamma_{k}^{\ms{(d)}} \sprod_{q=1}^{d} \mathrm{w}_{c_{q}k} .
\end{align*}

\begin{lemma}\label{lem:symmetric}
Each affinity tensor $\Lambdad$ defined element-wise in \Cref{eq:core-param1} is symmetric. 
\end{lemma}

\Cref{lem:symmetric} ensures our reparameterization of the affinity tensor preserves the symmetric invariance to permutations of classes. The next two lemmas provide context for our statement on identifiability. The first draws upon the separable non-negative matrix factorization literature~\cite{gillis2020nonnegative} to describe a unique matrix factorization. 
\begin{lemma}\label{lem:sep}
    For $\mathrm{W}$, as defined, the mapping $(\mathrm{\Theta}, \mathrm{W}) \to \mathrm{\Theta} \mathrm{W}$ is injective. 
\end{lemma}
The second gives conditions for which a CP tensor decomposition~\cite{hitchcock1927expression} is unique~\cite{sidiropoulos2000uniqueness}. 
\begin{lemma}[Uniqueness of CP]\label{lem:cp}
    The rank $K$ CP decomposition of the tensor $\Ad \in \mathbb{R}^{N \times \cdots \times N}$ given element-wise by 
    \begin{align}\Ad_{\mathbf{i}} = \sum_{k=1}^K \gamma_{k}^{\ms{(d)}} \sprod_{i \in \mathbf{i}} m_{i k} ,
    \end{align}
    is unique, up to permutation and scaling of the columns of the factor matrix $M = \mathrm{\Theta} \mathrm{W} \in \mathbb{R}^{N \times K}$ for $\gamma_{k}^{\ms{(d)}}, \sum_{i}m_{i k} > 0$ and $K \leq \frac{1}{2}(d(C-1) + 1)$. 
\end{lemma}
\noindent \textbf{Remark.} The decomposition is only unique, up to the permutation and scaling of the columns of $M$. \\
\textit{Justification.} The reconstructed tensor remains invariant to arbitrary permutations $\pi$ of the columns $[K]$, as $\Ad = \sum_{k=1}^K \tilde \gamma_{k} \bs{\tilde m}_{k}^{d}= \sum_{k=1}^K \tilde \gamma_{\pi(k)} \bs{\tilde m}_{\pi(k)}^{d}$ for any permutation $\pi$, i.e.,  the sum does not change. \\
Invariance under scaling is proved as follows. For any decomposition parameterized by $(M, \bs{\gamma})$, given by 
\begin{equation*}
\Ad = \sum_{k=1}^K \gamma_k \bs{m}_k^d = \sum_{k=1}^K \gamma_k \,(\bs{m}_k \circ \dots \circ \bs{m}_k),
\end{equation*}
 for arbitrary positive scaling constants $\psi_k > 0$,  there exists another decomposition parameterized by $(\tilde M, \bs{\tilde \gamma})$, with:
 \begin{align*}
 \tilde \gamma_k := \gamma_k \,(\psi_k)^d, \quad 
 \tilde m_{ik} := \frac{m_{ik}}{\psi_k} 
 \end{align*}
such that
\begin{align*}
\sum_{k=1}^K \tilde \gamma_k {\bs{\tilde m}}_k^{d} &= \sum_{k=1}^K \gamma_k (\psi_k)^d\, (\mathsmaller{\frac{\bs{m}_k}{\psi_k}})^{d}
= \sum_{k=1}^K \gamma_k \frac{\psi_k^d}{\psi_k^d}\bs{m}_k^d\\
&= \sum_{k=1}^K \gamma_k \bs{m}_k^d= \Ad.
\end{align*}

This result implies the uniqueness of the assortative model as defined in \Cref{thm:u1}.

 \begin{theorem}[\textit{Uniqueness of the strictly assortative model}]\label{thm:u1}
     Let $d \geq 3, \, \Lambdad_{c_1 \dots c_d} = \begin{cases}
         \gamma_{c}^{\ms{(d)}} & \text{if } c_j = c, \, \, \forall j \in [d]\\
         0 & \text{otherwise}
     \end{cases}$. Moreover, let $||\bs{\theta}_{c}||_1 = 1$ for each $c$. Then the decomposition given by 
     \begin{align} \label{eq:unique-HMT}
         \Mud_{i_1 \dots i_d} = \sum_{c=1}^C \gamma_{c}^{\ms{(d)}} \sprod_{r=1}^d \theta_{i_rc},
     \end{align}
     is unique up to the permutation of classes. 
 \end{theorem}
Note that this result is valid for $d\geq 3$, a result of~\cite{sidiropoulos2000uniqueness}, which holds for tensors of $d \geq 3$. When $d=2$, this model parameterizes a potentially non-unique, non-negative rank-$C$ matrix factorization $\Mud = \mathrm{\Theta} S \mathrm{\Theta}^\top$, where $S = \text{diag}(\bs{\gamma})$. Despite this,~\Cref{cor:id} shows that sharing parameters across $d \in \{2, \dots, D\}$ generically identifies the model.

We combine \Cref{lem:sep,lem:cp} to prove the following theorem, which states that under mild conditions (satisfied in all of our experiments) the parameters $(\mathrm{\Gamma}, \mathrm{\Theta}, \mathrm{W})$ uniquely determine $\Mud$ for some $d \leq D$ under the semi-assortative parameterization. 

\begin{theorem}[\textit{Uniqueness}]\label{thm:unique}
    Let the following hold: $d \geq 3, C < \text{Rank}(\Lambdad)=K \leq  \frac{1}{2}(d(C-1) + 1), \mathrm{W} = \begin{bmatrix}
        \mathrm{I}_C \mid \mathbf{w}_{c+1}, \dots, \mathbf{w}_K\\
    \end{bmatrix} \in \mathbb{R}^{C \times K}, \mathrm{\Theta} \in \mathbb{R}^{N \times C}, ||\mathrm{\Theta}||_{1,1} = \sum_{i=1}^N \sum_{c=1}^C \theta_{ic} =C,$ where $\mathrm{W}$ is column stochastic. Then the symmetric Tucker decomposition of $A \in \mathbb{R}^{N \times \dots \times N}$ with affinity tensor of rank $K$ and dimension $C \times \dots \times C$, given element-wise by ~\looseness=-1
    \begin{align}\label{eq:tucker}
        \Ad_{i_1\dots i_d} &= \sum_{c_1=1}^C \ldots \sum_{c_d = 1}^C \Lambdad_{\bs{c}} \sprod_{r=1}^d \theta_{i_r c_r}, \quad \Lambdad_{\bs{c}} = \sum_{k=1}^K \gamma_{k}^{\ms{(d)}} \sprod_{c \in \bs{c}}\mathrm{w}_{ck}
    \end{align}
    and parameterized by $\left(\bs{\gamma}, \mathrm{W}, \mathrm{\Theta} \right)$ is unique, up to the permutation of classes $[C]$ and communities $[K]$.
\end{theorem}

The condition that $K$ is sufficiently small is not satisfied for all $d \in \{2, \dots, D\},$ but it is satisfied for some $d \leq D$ in all of our experiments (the bound is increasing in $d$ and $C$), which is sufficient to ensure our model is identifiable. In our experiments, the tightest bound occurs in the hospital setting, where the upper bound is tight: $\frac{1}{2}(D(C -1) + 1) = \frac{1}{2}(5(2 -1) + 1) = 3 = K$. ~\looseness=-1 

\begin{definition}[\textit{Generic identifiability}]
We say that  the model defined by \Cref{eq:mu,eq:core-param1} is \textit{generically identifiable} if
\begin{align*}
   \mathbb{P}(\Aall \mid \mathrm{\Gamma}, \mathrm{W}, \mathrm{\Theta}) = \mathbb{P}(\Aall \mid \tilde {\mathrm{\Gamma}}, \tilde {\mathrm{W}}, \tilde {\mathrm{\Theta}})
   \end{align*} if and only if 
   \begin{align*}
   (\mathrm{\Gamma}, \mathrm{W}, \mathrm{\Theta}) = (\tilde {\mathrm{\Gamma}}, \tilde {\mathrm{W}}, \tilde {\mathrm{\Theta}}),
\end{align*}
almost surely over the parameter space $\mathbb{R}_{>0}^{D - 1 \times K} \times (\mathrm{\Delta}^{C-1})^{K-C} \times (\mathrm{\Delta}^{N-1})^C$ up to permutations of latent classes $[C]$ and communities $[K]$, where $\mathrm{\Delta}^{p-1} := \{\mathbf{x} \in \mathbb{R}^p: \mathrm{x}_i \geq 0, ||\mathbf{x}||_1 = 1\}$ is the $p$-dimensional simplex in $\mathbb{R}^p$.~\looseness=-1
\end{definition}

\begin{corollary}[Identifiability]\label{cor:id}
    The model over the adjacency tensors $\Aall = \{A^{\ms{(2)}}, \dots, A^{\ms{(D)}}\}$ defined by \Cref{eq:mu,eq:core-param1} is generically identifiable. 
\end{corollary}

\textbf{Remark.} Similarly, the model defined by the strictly assortative model is identifiable. Often, the separability condition $\mathrm{W} = [\mathrm{I}_C \mid \mathbf{w}_{c+1}, \dots,\mathbf{w}_{K} ]$ can be prohibitive, requiring optimization methods that identify $C$ columns of $\mathrm{W}$ which correspond to the identity matrix $\mathrm{I}_C$, where $\mathrm{w}_{cc'} = 0$ unless $c = c'$, in which case $\mathrm{w}_{cc'}=1$, i.e.,  columns are one-hot encoded:
\begin{equation*}
[\mathbf{w}_{1}, \dots, \mathbf{w}_{C}] = \textrm{I}_C.
\end{equation*}
Often, these columns correspond to observed variables or features in the data and cannot be assigned arbitrarily. However, in our setting, both dimensions of $\mathrm{W}$ are latent, and so we arbitrarily assign the first $C$ columns as pivots at initialization and learn the community structure accordingly.

\Cref{lem:dis-unique} shows the omniassortative model may be alternatively parameterized as an extension of the semi-assortative model to include negative scaling constants.

\begin{lemma} \label{lem:dis-unique}
    Let $\Mud$ be the $N \times \dots \times N$ tensor defined element-wise by the omniassortative model as defined according to~\Cref{eq:omni-core}.
Then the reconstructed tensor $\Mud$ has the CP decomposition representation
\begin{align*}
    \Mud_{\mathbf{i}} = \sum_{k=1}^K \tilde \gamma_{k}^{\ms{(d)}} \sprod_{i \in \mathbf{i}} \bs{\theta}_{i}^\top \mathbf{w}_k,
\end{align*}
where $\tilde \gamma_{k}^{\ms{(d)}} = \begin{cases} 
\gamma_{k}^{\ms{(d)}} - \sum_{k'=C+1}^K \gamma_{k'}^{\ms{(d)}} \mathrm{w}_{kk'}^d & \text{ if } k \in [C]\\
\gamma_{k}^{\ms{(d)}} & \text{ otherwise.}\end{cases}$
\end{lemma}

\Cref{lem:dis-unique} lets us apply the results of \Cref{thm:unique} to the omniassortative model.

\begin{corollary}\label{cor:du}
    Under the same set of assumptions described in \Cref{thm:unique}, the decomposition defined by the omniassortative model is unique, up to permutation and scaling. Let $||\mathbf{w}_k||_1 = 1$ and $||\mathrm{\Theta}||_{1,1} = C$. Then the decomposition is unique up to permutation of classes $[C]$ and communities $[K]$.   
\end{corollary}

By \Cref{thm:unique}, $\tilde {\mathrm{\Gamma}}$ and $\mathrm{\Theta}$, and $\mathrm{W}$ are uniquely determined. 
For $c \in [C]$, we may solve for $\gamma_{c}^{\ms{(d)}} = \tilde \gamma_{c}^{\ms{(d)}} + \sum_{k=C+1}^K \tilde \gamma_{k}^{\ms{(d)}} \mathrm{w}_{ck}^d$ and for $k > C$, $\gamma_{k}^{\ms{(d)}} = \tilde \gamma_{k}^{\ms{(d)}}$. Leveraging this result, we claim the omniassortative model is generically identifiable. 

\subsection*{Proofs}\label{sec:proofs}
\textbf{Proof of inequality~\ref{lemma:prop}}.
\begin{proof}
    The proof relies on Jensen's inequality, which states that for convex function $f: \mathbb{R} \to \mathbb{R}$
    \begin{equation*}
    f(\mathsmaller{\frac{1}{n} \sum_{i=1}^n x_i}) \leq \frac{1}{n} \sum_{i=1}^nf(x_i).
    \end{equation*}
     Letting $f(x) = x^d$, which is convex for $x \geq 0$, implies
    \begin{align*}
        \left(\frac{1}{C} \sum_{c=1}^C \mathrm{w}_{ck}\right)^d = f\left(\tfrac{1}{C}\sum_{c=1}^C \mathrm{w}_{ck}\right) &\leq \frac{1}{C}\sum_{c=1}^C f(\mathrm{w}_{ck}) = \frac{1}{C} \sum_{c=1}^C \mathrm{w}_{ck}^d.
    \end{align*}
    Consider an element $\Lambdad_{c \dots c} = \sum_{k=1}^K \gamma_{k}^{\ms{(d)}} \sprod_{r=1}^d \mathrm{w}_{c_rk} = \sum_{k=1}^K \gamma_{k}^{\ms{(d)}} \mathrm{w}_{ck}^d$ on the diagonal of $\Lambdad$. 
    Then 
    \begin{align*}
    \frac{1}{C} ||\text{diag}(\Lambdad)||_{1,1} &= \frac{1}{C}\sum_{c=1}^C \Lambdad_{c \dots c}\\
    &= \frac{1}{C}\sum_{c=1}^C \sum_{k=1}^K \gamma_{k}^{\ms{(d)}} \sprod_{r=1}^d \mathrm{w}_{c_rk} &&\text{(def. of $\Lambdad_{c \dots c}$)}\\
    &= \sum_{k=1}^K \gamma_{k}^{\ms{(d)}} \frac{1}{C} \sum_{c=1}^C \mathrm{w}_{ck}^d &&  (c_r = c, \,  \forall r \implies \sprod_{r=1}^d \mathrm{w}_{c_rk} = \mathrm{w}_{ck}^d) \\
    & \geq \sum_{k=1}^K \gamma_{k}^{\ms{(d)}} \frac{1}{C^d} \left(\sum_{c=1}^C \mathrm{w}_{ck}\right)^d  && \text{(Jensen's)}\\
    &= \frac{1}{C^d} \sum_{k=1}^K \gamma_{k}^{\ms{(d)}} \sum_{c_1=1}^C \ldots \sum_{c_d=1}^C  \sprod_{r=1}^d \mathrm{w}_{c_rk} \\
    &=  \frac{1}{C^d} \sum_{c_1=1}^C \ldots \sum_{c_d=1}^C  \sum_{k=1}^K \gamma_{k}^{\ms{(d)}} \sprod_{r=1}^d \mathrm{w}_{c_rk}\\
    &= \frac{1}{C^d} \sum_{c_1=1}^C \ldots \sum_{c_d=1}^C  \Lambdad_{c_1 \dots c_d} && (\Lambdad_{c_1 \dots c_d} = \sum_{k=1}^K \gamma_{k}^{\ms{(d)}} \sprod_{r=1}^d \mathrm{w}_{c_rk})\\
    &= \frac{1}{C^d} ||\Lambdad||_{1,}.
    \end{align*}
    Multiplying both sides by $\frac{C}{||\Lambdad||_1}$ yields
        \begin{align}
                    \frac{||\text{diag}(\Lambdad)||_{1,1}}{||\Lambdad||_{1,1}}  \geq \frac{1}{C^{d-1}}.
        \end{align}
\end{proof}

\textbf{Proof of \Cref{lem:symmetric}}.
\begin{proof}  Consider an element of the $d$th affinity tensor $\Lambdad_{c_1 \dots c_d} := \Lambdad_{\bs{c}}$ and $\Lambdad_{\pi(\bs{c})}$, where $\pi(\bs{c}) := \left(\pi(c_1),\dots, \pi(c_d)\right)$ is an arbitrary permutation of the indices $\bs{c} := (c_1, \dots, c_d)$.~\looseness=-1
Then $\Lambdad_{\bs{c}} = \Lambdad_{\pi(\bs{c})}$:
\begin{align*}
    \Lambdad_{\bs{c}} &= \sum_{k=1}^K \gamma_{k}^{\ms{(d)}} \sprod_{r=1}^d \mathrm{w}_{c_rk}\\
    &= \sum_{k=1}^K \gamma_{k}^{\ms{(d)}} \left(\mathrm{w}_{c_1 k} \, \mathrm{w}_{c_2 k} \dots  \mathrm{w}_{c_d k}\right)\\
    &= \sum_{k=1}^K \gamma_{k}^{\ms{(d)}} \left(\mathrm{w}_{\pi(c_1) k} \, \mathrm{w}_{\pi(c_2) k} \, \dots \, \mathrm{w}_{\pi(c_d) k}\right)\\
    &= \sum_{k=1}^K \gamma_{k}^{\ms{(d)}} \sprod_{r=1}^d \mathrm{w}_{ \pi(\bs{c}_r) k} = \Lambdad_{\pi(\bs{c})} .
\end{align*}
\end{proof}

\textbf{Proof of \Cref{lem:sep}.}
\begin{proof}
    Let  $\mathrm{\Theta} \mathrm{W} = M = \tilde {\mathrm{\Theta}} \tilde {\mathrm{W}}$. Then 
    \begin{align*}
    [\mathrm{\Theta} \mid \mathrm{\Theta} [\mathbf{w}_{c+1}, \dots, \mathbf{w}_K]] = \mathrm{\Theta} [\mathrm{I}_C \mid \mathbf{w}_{c+1}, \dots, \mathbf{w}_K] = \mathrm{\Theta} \mathrm{W} = M = \tilde {\mathrm{\Theta}} \tilde {\mathrm{W}} = \tilde {\mathrm{\Theta}} [\mathrm{I}_C \mid \tilde{\mathbf{w}}_{c+1}, \dots, \tilde{\mathbf{w}}_{K}] = [\tilde {\mathrm{\Theta}} \mid \tilde {\mathrm{\Theta}} [\tilde{\mathbf{w}}_{c+1}, \dots, \tilde{\mathbf{w}}_{K}]].
    \end{align*}
    That is, $\mathrm{\Theta} = \tilde {\mathrm{\Theta}}$ and
    $\mathrm{W} = (\mathrm{\Theta}^{\dagger}\mathrm{\Theta}) \mathrm{W} = \mathrm{\Theta}^{\dagger}(\mathrm{\Theta} \mathrm{W}) = \mathrm{\Theta}^\dagger M = \tilde {\mathrm{\Theta}}^{\dagger} M = \tilde {\mathrm{\Theta}}^{\dagger} \tilde {\mathrm{\Theta}} \tilde {\mathrm{W}} = \tilde {\mathrm{W}}$ (assuming $\mathrm{\Theta}$ is full rank), where $\mathrm{\Theta}^{\dagger}$ denotes the Moore-Penrose pseudoinverse~\cite{ben2006generalized} of $\mathrm{\Theta}$. Together, $(\mathrm{\Theta}, \mathrm{W}) = (\tilde {\mathrm{\Theta}}, \tilde {\mathrm{W}}).$
\end{proof}

\textbf{Proof of \Cref{lem:cp}.}
\begin{proof}
We begin with the definition of \textit{Kruskal rank}. 
\begin{definition}(Kruskal rank)
    The \textbf{Kruskal rank} $\mathcal{K}_B$ of a matrix $B$ is the maximal $\mathcal{K}_B$ such that any $\mathcal{K}_B$ columns of $B$ are linearly independent. 
\end{definition}
The Kruskal rank of $M = \mathrm{\Theta} \mathrm{W}$ is $\mathcal{K}_M = C$ for all $\mathrm{\Theta}, \mathrm{W} \in (\mathrm{\Delta}^{N-1})^C \times I_C \times (\mathrm{\Delta}^{C-1})^{K-C}$ except for the measure-zero subset where the subsets of columns of $\mathrm{W}$ are linearly dependent or $\mathrm{\Theta}$ is rank deficient. 

The result is a corollary of \cite{sidiropoulos2000uniqueness}, who extend Kruskal's theorem~\cite{kruskal1989rank} to show the decomposition of the rank $K$ tensor $\Ad \in \mathbb{R}^{N_1 \times \dots \times N_d}$ is unique (up to permutation and scaling) if $\sum_{i=1}^D {\mathcal{K}_{B_i}} \geq 2K + (D - 1),$ where $\mathcal{K}_{B_i}$ is the Kruskal rank of the $i$th factor matrix $B_i \in \mathbb{R}^{N_i \times K}$. This result holds for tensors not restricted to be non-negative or symmetric. We apply the result to our setting, where $C = \mathcal{K}_i$ for each $i \in [D]$, which yields $DC \geq 2K + (D-1)$. Isolating $K$ yields the sufficient condition 
$K \leq \frac{1}{2}(D(C-1) + 1)$.
\end{proof}

\textbf{Proof of \Cref{thm:u1}.}
   \begin{proof}
     By \Cref{lem:cp}, \Cref{eq:unique-HMT} is unique, up to permutation and scaling of the columns of $\mathrm{\Theta}$. That is, if $(\mathrm{\Gamma}, \mathrm{\Theta})$ and $(\tilde {\mathrm{\Gamma}}, \tilde {\mathrm{\Theta}})$ parameterize \Cref{eq:unique-HMT}, then $\tilde \gamma_{c}^{\ms{(d)}} = \frac{\tilde \gamma_{d\pi(c)}}{\psi_{c}^d}$ and $\bs{\tilde \theta}_c = \psi_{c} \bs{\theta}_{\pi(c)}$ for some permutation $\pi$ and scalars $\psi_c > 0$. The $\ell_1$ constraint on each column of $\mathrm{\Theta}$ implies that $\psi_c \!=\! 1$ for all $c \in [C]$, that is, 
     \begin{align*}
     1 = || \bs{\tilde \theta}_c||_1 = ||\psi_{c}\bs{ \theta}_{\pi(c)}||_1 = \psi_c||\bs{\theta}_{\pi(c)}||_1 = \psi_c,
     \end{align*}
     and so $\bs{\tilde \theta}_c = \bs{\theta}_{\pi(c)}$ for each $c$.
\end{proof}
\textbf{Proof of \Cref{thm:unique}.}
\begin{proof}
    We plug in $\Lambdad_{c_1\dots c_d} = \sum_{k=1}^K \gamma_{k}^{\ms{(d)}}\sprod_{r=1}^d \mathrm{w}_{c_rk}$ to re-express \Cref{eq:tucker} as 
    \begin{align}
        \Ad_{i_1\dots i_d} &= \sum_{c_1=1}^C \ldots \sum_{c_d=1}^C   \sum_{k=1}^K \gamma_{k}^{\ms{(d)}} \sprod_{r=1}^d \mathrm{w}_{c_rk} \sprod_{r=1}^d \theta_{i_r c_r} && (\text{plugging in }\Lambdad_{c_1\dots c_d} = \sum_{k=1}^K \gamma_{k}^{\ms{(d)}} \sprod_{r=1}^d \mathrm{w}_{c_rk})\\
        &= \sum_{k=1}^K  \gamma_{k}^{\ms{(d)}} \sum_{c_1=1}^C \ldots \sum_{c_d=1}^C  \sprod_{r=1}^d \mathrm{w}_{c_rk} \sprod_{r=1}^d \theta_{i_r c_r} && \text{(swap order of summation)}\\
        &= \sum_{k=1}^K   \gamma_{k}^{\ms{(d)}} \sprod_{r=1}^d \left(\sum_{c_r=1}^C \mathrm{w}_{c_rk} \theta_{i_d c_r}\right) && \text{(push sum into product)}\\
        &= \sum_{k=1}^K  \gamma_{k}^{\ms{(d)}} \sprod_{r=1}^d \vm_{i_r}^\top\mathbf{w}_k && (\text{simplify}) \\
        &= \sum_{k=1}^K \gamma_{k}^{\ms{(d)}} \sprod_{r=1}^d m_{i_r k} && m_{i_r k} :=  \vm_{i_r}^\top\mathbf{w}_k.\label{cp:representation}
    \end{align}
    We make two observations, which help us finish the proof.
    \begin{itemize}
        \item \Cref{cp:representation} parametrizes a (symmetric) rank $K$ CP decomposition with factor matrix $M = \mathrm{\Theta} \mathrm{W} \in \mathbb{R}^{N \times K}$. \Cref{lem:cp} implies this decomposition is unique, up to permutations of the columns  of $M$ and scaling constants $\bs{\psi}$.
        \item  By \Cref{lem:sep}, $\mathrm{\Theta}$ and $\mathrm{W}$ uniquely construct $M = \mathrm{\Theta} \mathrm{W}$.
    \end{itemize}
     
        Suppose that two distinct parameterizations $(\mathrm{\Gamma}, \mathrm{\Theta}, \mathrm{W})$ and $(\tilde {\mathrm{\Gamma}}, \tilde {\mathrm{\Theta}}, \tilde {\mathrm{W}})$ reconstruct $A$. Then \Cref{cp:representation} implies that 
        \begin{equation*}
        \mathrm{\Theta} \mathrm{W} = M, \tilde {\mathrm{\Theta}} \tilde {\mathrm{W}} = \tilde {M},
        \end{equation*}
        where $M = \tilde M$, up to column permutation and scaling. We show that under the constraints $\mathrm{W} = \begin{bmatrix}\mathrm{I}_C \mid \mathbf{w}_{c+1}, \dots, \mathbf{w}_K\end{bmatrix}$, $||\mathrm{\Theta}||_{1,1} = C$, then $(\mathrm{\Theta}, \mathrm{W}) = (\tilde {\mathrm{\Theta}}, \tilde {\mathrm{W}})$ must hold. 

        By \Cref{cp:representation}, each column of $\tilde M$ and scalar $\tilde \psi_{dk}$ can be expressed as 
        \begin{align*}
             \bs{\tilde m}_k = \psi_k \bs{m}_{\pi(k)}, \hspace{1em} \tilde \gamma_{k}^{\ms{(d)}} = \frac{\gamma_{d\pi(k)}}{\psi_{k}^d}.
        \end{align*}
Then 
    \begin{align*}
    \sum_{c=1}^C \tilde \theta_{ic} \tilde {\mathrm{w}}_{ck} = \bs{\tilde m}_{ik} = \psi_k \bs{m}_{i \pi(k)} = \psi_k  \sum_{c=1}^C \theta_{ic} \mathrm{w}_{c \pi(k) }. 
    \end{align*}
    Since $\sum_{c=1}^C \mathrm{w}_{c \pi(k)} = \sum_{c=1}^C \mathrm{w}_{c k} = 1$ it must hold that $\bs{\tilde m}_k = \psi_{k} \mathrm{\Theta} \mathbf{w}_{\pi(k)}$, i.e.,  $\mathrm{W} = \tilde {\mathrm{W}} \mathrm{\Pi}_K$, for some permutation $\mathrm{\Pi}_K$. Then $\tilde {\mathrm{\Theta}} = \psi_k \mathrm{\Theta}$ for all $k$, which implies that $\psi_k = \psi$ for all $k$. But then $C = ||\tilde {\mathrm{\Theta}}||_{1,1} = ||\psi \mathrm{\Theta}||_{1,1} = \psi ||\mathrm{\Theta}||_{1,1} = \psi C$ and so $\psi = 1$. But then $\mathrm{\Theta} = \mathrm{\Pi}_C \tilde {\mathrm{\Theta}}$, for some permutation $\mathrm{\Pi}_C$, and so the two parameterizations are equivalent. 
\end{proof}

\textbf{Proof of \Cref{cor:id}.}
\begin{proof} 
    Suppose $\mathbb{P}(\Aall \mid \mathrm{\Gamma}, \mathrm{W}, \mathrm{\Theta}) = \mathbb{P}(\Aall \mid \tilde {\mathrm{\Gamma}}, \tilde {\mathrm{W}}, \tilde {\mathrm{\Theta}})$. Then for each $d \in \{2, \dots, D\}$, $i_1 < \dots < i_d$, $\Mud_{i_1 \dots i_d} = \tilde \Capmu^{\ms{(d)}}_{i_1 \dots i_d}$. For $\bs{m}_k = \mathrm{\Theta} \mathbf{w}_k \in \mathbb{R}^{N}_{>0}$, 
    \begin{align*}
        \Mud_{i_1 \dots i_d} = \sum_{k=1}^K \gamma_{k}^{\ms{(d)}} \bs{m}_k \circ \dots \circ \bs{m}_k, \quad \tilde \Capmu^{\ms{(d)}}_{i_1 \dots i_d} = \sum_{k=1}^K \tilde \gamma_{k}^{\ms{(d)}} {\bs{\tilde m}}_k \circ \dots \circ {\bs{\tilde m}}_k
    \end{align*}
    the ``residual" tensor $R^{\ms{(d)}} := \tilde \Capmu^{\ms{(d)}} - \Mud$ is zero on the indices $i_1 \neq \dots \neq i_d$, and potentially nonzero elsewhere. 
    
    Thus, $R^{\ms{(d)}}$ contains (at least) ${N \choose d}$ entries that are exactly zero, one for each $i_1 < \dots < i_d$.  Suppose that $R^{\ms{(d)}}$ is not the $0$ tensor, i.e.,  $R^{\ms{(d)}} \neq 0$. By assumption, $\text{Rank}(\Mud) = K = \text{Rank}(\tilde \Capmu^{\ms{(d)}}) = \text{Rank}(\Mud + R^{\ms{(d)}}),$ which implies that $R^{\ms{(d)}}$ lies in the span of the rank-one tensors $(\bs{m}_k \circ \dots \circ \bs{m}_k)_{k=1}^K$, i.e.,  $R^{\ms{(d)}} = \sum_{k=1}^K r_k \bs{m}_k \circ \dots \circ \bs{m}_k$ for some constants $r_k \in \mathbb{R}$. Since each entry $m_{ik} > 0$ and there are $K < {N \choose d}$ rank-$1$ tensors, $R^{\ms{(d)}}$ lies outside the span of the $K$ rank-one tensors almost surely. This implies that for $d \geq 3$, $\Mud\!=\!\tilde \Capmu^{\ms{(d)}}$. 
    
    Consider the case where $R^{\ms{(d)}} = 0$, i.e.,  $\Mud = \tilde \Capmu^{\ms{(d)}}$. By \Cref{thm:unique}, for large enough $d$, $\Mud$ is uniquely defined, with factor matrix $M = \mathrm{\Theta} \mathrm{W}$, (which is uniquely defined by $\mathrm{\Theta}$ and $\mathrm{W}$), up to permutation. Therefore, it must be that $\mathrm{\Theta} = \tilde {\mathrm{\Theta}}$ and $\mathrm{W} = \tilde{\mathrm{W}}$. We remove scaling ambiguity of the constants $\gamma_{k}^{\ms{(d)}}$ by requiring $||\mathbf{w}_k||_1 = ||\bs{\theta}_c||_1 = 1$, which implies that $||\bs{m}_k||_1 = \sum_{i=1}^N \sum_{c=1}^C \theta_{ic} \mathrm{w}_{ck} = 1$. 

Let $\bs{\gamma}_d = (\gamma_{k}^{\ms{(d)}})_{k=1}^K$. Consider the matrix setting, $d=2$ and suppose $\bs{\gamma}_{2} \neq \bs{\tilde \gamma}_2$. The higher-order parameterization  ($d \geq 3$) requires that $\mathrm{\Theta}, \mathrm{W}$ are unique and together they define $M = \mathrm{\Theta} \mathrm{W}$. Suppose that $\Capmu_{ij}^{\ms{(2)}} = \tilde \Capmu_{ij}^{\ms{(2)}} $ for all $i \neq j$. Similar to above, $\Mud_{ii} \neq \tilde \Capmu^{\ms{(d)}}_{ii}$ occurs with probability $0$. Thus, suppose $\Capmu^{\ms{(2)}} = \tilde \Capmu^{\ms{(2)}}$. Then $\mathrm{\Theta} \mathrm{W} S \mathrm{W}^\top \mathrm{\Theta}^\top = \mathrm{\Theta} \mathrm{W} \tilde S \mathrm{W}^\top \mathrm{\Theta}^\top$, and taking the pseudo-inverse of $\mathrm{\Theta}$ implies that $\mathrm{W} S \mathrm{W}^\top = \mathrm{W} \tilde S \mathrm{W}^\top$. Since $\mathrm{W}$ is separable with $l_1$-norm constrained columns, it follows that $S = \tilde S$~\cite{gillis2020nonnegative}. Similar for $d \geq 3$, the constants are uniquely defined almost surely.  ~\looseness=-1   

For $(\mathrm{\Gamma}, \mathrm{\Theta}, \mathrm{W}) = (\tilde {\mathrm{\Gamma}}, \tilde {\mathrm{\Theta}}, \tilde {\mathrm{W}})$, the tensor reconstruction given by \Cref{eq:mu,eq:core-param1} is well-defined, implying that 
\begin{equation*}
\mathbb{P}(\Aall \mid \mathrm{\Gamma}, \mathrm{W}, \mathrm{\Theta}) = \mathbb{P}(\Aall \mid \tilde {\mathrm{\Gamma}}, \tilde {\mathrm{W}}, \tilde {\mathrm{\Theta}}).
\end{equation*}
\end{proof}

\textbf{Proof of \Cref{lem:dis-unique}.}
\begin{proof}
    Noting that for $C < k \leq K$, each sum over $c_1, \dots, c_D$ may be written by computing the sum over all terms (ignoring the indicator), and then subtracting the terms s.t. $1(\exists c_i \neq c_j) = 0$, which occurs for $(c_1, \dots, c_d) = (c, \dots , c)$ for $c \in [C]$. For a given $k$, $C < k \leq K$, 
    \begin{equation*}
    \sum_{c_1=1}^C \ldots \sum_{c_d=1}^C  1(\exists c_i \neq c_j)\gamma_{k}^{\ms{(d)}} \sprod_{r=1}^d \mathrm{w}_{c_rk}\theta_{i_r c_r} = \gamma_{k}^{\ms{(d)}} \left(\sum_{c_1=1}^C \ldots \sum_{c_d=1}^C   \sprod_{r=1}^d \mathrm{w}_{c_rk}\theta_{i_r c_r} -  \sum_{c=1}^C  \mathrm{w}_{c k}^d \sprod_{r=1}^d \theta_{i_r c_r} \right).
    \end{equation*}
    We re-express $\Mud_{i_1 \dots i_d}$ as 
    \begin{align}
    \Mud_{i_1 \dots i_d} &= \sum_{c_1=1}^C \ldots \sum_{c_d=1}^C  \sum_{k=1}^C \gamma_{k}^{\ms{(d)}} \sprod_{r=1}^d\mathrm{w}_{c_rk} \theta_{i_r c_r} + \sum_{c_1=1}^C \ldots \sum_{c_d=1}^C  1(\exists c_i \neq c_j) \sum_{k = C+1}^K \gamma_{k}^{\ms{(d)}} \sprod_{r=1}^d \mathrm{w}_{c_rk} \theta_{i_r c_r} \\
    &= \sum_{c=1}^C \gamma_{c}^{\ms{(d)}} \mathrm{w}_{cc}^d  \sprod_{r=1}^d \theta_{i_r c} + \sum_{k=C+1}^K \gamma_{k}^{\ms{(d)}}\left(\sum_{c_1=1}^C \ldots \sum_{c_d=1}^C   \sprod_{r=1}^d \mathrm{w}_{c_rk} \theta_{i_r c_r} - \sum_{c=1}^C \mathrm{w}_{ck}^d \sprod_{r=1}^d \theta_{i_r c} \right)\\
    &= \sum_{c=1}^C \gamma_{c}^{\ms{(d)}}  \sprod_{r=1}^d \theta_{i_r c} + \sum_{k=C+1}^K \gamma_{k}^{\ms{(d)}}\left(\sprod_{r=1}^d \left(\sum_{c=1}^C \mathrm{w}_{ck} \theta_{i_r c}\right) - \sum_{c=1}^C \mathrm{w}_{ck}^d \sprod_{r=1}^d \theta_{i_r c} \right)\\
    &= \sum_{c=1}^C \gamma_{c}^{\ms{(d)}}  \sprod_{r=1}^d \theta_{i_r c} + \sum_{k=C+1}^K \gamma_{k}^{\ms{(d)}}\left(\sprod_{r=1}^d m_{i_r k} - \sum_{c=1}^C \mathrm{w}_{ck}^d \sprod_{r=1}^d \theta_{i_r c} \right).\label{eq:d_decomp}
    \end{align}
    Combining like terms in \Cref{eq:d_decomp}, we obtain
    \begin{align*}
        \Mud_{i_1 \dots i_d} &= \sum_{c=1}^C \gamma_{c}^{\ms{(d)}} \sprod_{r=1}^d m_{i_r c} - \sum_{c=1}^C \sum_{k=C+1}^K \gamma_{k}^{\ms{(d)}} \mathrm{w}_{ck}^d \sprod_{r=1}^d m_{i_d c} + \sum_{k=C+1}^K \gamma_{k}^{\ms{(d)}} \sprod_{r=1}^d m_{i_r k}\\
        &= \sum_{c=1}^C \underbrace{\left(\gamma_{c}^{\ms{(d)}} - \sum_{k=C+1}^K \gamma_{k}^{\ms{(d)}}\mathrm{w}_{ck}^d\right)}_{:=\tilde \gamma_{c}^{\ms{(d)}}}\sprod_{r=1}^d m_{i_r c} + \sum_{k=C+1}^K \underbrace{\gamma_{k}^{\ms{(d)}}}_{:=\tilde \gamma_{k}^{\ms{(d)}}} \sprod_{r=1}^d m_{i_r k}\\
        &= \sum_{k=1}^K \tilde \gamma_{k}^{\ms{(d)}} \sprod_{r=1}^d m_{i_r k}.
    \end{align*}
\end{proof}

\section*{Supplementary Note 5: Detailed derivations}

In Supplementary Note 1 we give the updates and the general framework for how we derive them. Here we explicitly derive them in more detail. Specifically, we give full derivations of the optimal updates for the semi-assortative model. The derivations of the fully omniassortative updates are similar to the derivations here---the derivations for the strictly assortative derivations are given by \cite{contisciani2022inference}. Abusing notation, we write $\mathcal{Q}$ to mean $\mathcal{Q}(\Aall, \mathrm{\Gamma}, \mathrm{W}, \mathrm{\Theta})$. 

The evidence lower bound is given by:
\begin{align}
    \mathcal{Q} &= - \sum_{d=2}^D \sum_{k=1}^K \gamma_{k}^{\ms{(d)}} \phi_{k}^{\ms{(d)}} + \sum_{d=2}^D \sum_{k=1}^K \sum_{\mathbf{i} \in \Omegad} \Exp\left[\Ad_{\mathbf{i}k} \mid \Aall\right] \log (\gamma_{k}^{\ms{(d)}}) + \sum_{d=2}^D \sum_{k=1}^K \sum_{\mathbf{i} \in \Omegad}  \sum_{r=1}^d \sum_{c_r=1}^C \Exp \left[\Ad_{\mathbf{i}i_r c_r k} \mid \Aall\right] \log (\theta_{i_r c_r}\mathrm{w}_{c_rk}) \\
    &= - \sum_{d=2}^D \sum_{k=1}^K \gamma_{k}^{\ms{(d)}} \phi_{k}^{\ms{(d)}} + \sum_{d=2}^D \sum_{k=1}^K \sum_{\mathbf{i} \in \Omegad} \Exp\left[\Ad_{\mathbf{i}k} \mid \Aall\right] \log (\gamma_{k}^{\ms{(d)}}) +  \sum_{k=1}^K  \sum_{i=1}^N 1(i \in \mathbf{i})\sum_{c=1}^C \sum_{d=2}^D \sum_{\mathbf{i} \in \Omegad} \Exp\left[\Ad_{\mathbf{i}ick} \mid \Aall\right] \log (\theta_{i c }\mathrm{w}_{kc }) \\
    &= - \sum_{d=2}^D \sum_{k=1}^K \gamma_{k}^{\ms{(d)}} \phi_{k}^{\ms{(d)}} + \sum_{d=2}^D \sum_{k=1}^K \sum_{\mathbf{i} \in \Omegad} \Exp\left[\Ad_{\mathbf{i}k} \mid \Aall\right] \log (\gamma_{k}^{\ms{(d)}}) +  \sum_{k=1}^K  \sum_{i=1}^N \sum_{c=1}^C \Exp\left[\Ad_{ick} \mid \Aall\right] \log (\theta_{i c }\mathrm{w}_{kc })\\
    &= - \sum_{d=2}^D \sum_{k=1}^K \gamma_{k}^{\ms{(d)}} \phi_{k}^{\ms{(d)}} + \sum_{d=2}^D \sum_{k=1}^K \sum_{\mathbf{i} \in \Omegad} \Exp\left[\Ad_{\mathbf{i}k} \mid \Aall\right] \log (\gamma_{k}^{\ms{(d)}}) +  \sum_{i=1}^N \sum_{c=1}^C \log (\theta_{ic}) \sum_{k=1}^K \Exp\left[\Ad_{ick} \mid \Aall\right]  \\
    &+  \sum_{k=1}^K \sum_{c=1}^C  \log(\mathrm{w}_{ck})\sum_{i=1}^N \Exp\left[\Ad_{ick} \mid \Aall\right] .\label{eq:ELBO}
\end{align} 
\noindent \textbf{M-step.}

\textbf{Update to }$\bs{\theta_{ic}}.$ To update $\theta_{ic}$, we set the partial derivative
$
\frac{\partial \mathcal{Q}}{\partial \theta_{ic}} = 0$
and solve for $\theta_{ic}$. By definition, 
\begin{align*}
    \frac{\partial m_{ik}}{\partial \theta_{ic}} = \mathrm{w}_{ck}, \hspace{1em} \frac{\partial \phi_{k}^{\ms{(d)}}}{\partial m_{i k}} = \sum_{\mathbf{i} \in \Omegad, i \in \mathbf{i}} \sprod_{i_r \neq i} m_{i_r k} := \bar \phi_{ik}^{\ms{(d-1)}}.
\end{align*}
The partial derivative is given by:
\begin{align*}
    \frac{\partial \mathcal{Q}}{\partial \theta_{ic}} &= - \sum_{d=2}^D \sum_{k=1}^K \gamma_{k}^{\ms{(d)}} \left(\frac{\partial \phi_{k}^{\ms{(d)}}}{\partial m_{i k}} \cdot \frac{\partial m_{ik}}{\partial \theta_{ic}}\right) + \frac{1}{\theta_{i c}} \sum_{k=1}^K \Exp[\Ad_{ick} \mid \Aall]\\
    &= \sum_{d=2}^D \sum_{k=1}^K \gamma_{k}^{\ms{(d)}} \bar \phi_{ik}^{\ms{(d-1)}} \mathrm{w}_{ck}  + \frac{1}{\theta_{i c}} \sum_{k=1}^K \Exp[\Ad_{ick} \mid \Aall].
\end{align*}
Setting $\frac{\partial \mathcal{Q}}{\partial \theta_{ic}} = 0$ and solving for $\theta_{ic}$ yields 
\begin{align*}
    \theta_{ic} = \frac{\sum_{k=1}^K \Exp[\Ad_{ick} \mid \Aall]}{\sum_{d=2}^D \sum_{k=1}^K \gamma_{k}^{\ms{(d)}} \bar \phi_{ik}^{\ms{(d-1)}} \mathrm{w}_{ck}}.
\end{align*}

\textbf{Update to $\bs{\gamma_{k}^{\ms{(d)}}}$.} The partial derivative $\frac{\partial Q}{\partial \gamma_{k}^{\ms{(d)}}}$ is given by:
\begin{align*}
    \frac{\partial Q}{\partial \gamma_{k}^{\ms{(d)}}} = - \phi_{k}^{\ms{(d)}} + \frac{1}{\gamma_{k}^{\ms{(d)}}}\sum_{\mathbf{i} \in \Omegad} \Exp\left[\Ad_{\mathbf{i} k} \mid \Aall \right].
\end{align*}
Setting $\frac{\partial Q}{\partial \gamma_{k}^{\ms{(d)}}} = 0$ and solving for $\gamma_{k}^{\ms{(d)}}$ yields:
\begin{align*}
    \gamma_{k}^{\ms{(d)}} = \frac{\sum_{\mathbf{i} \in \Omegad} \Exp\left[\Ad_{\mathbf{i} k} \mid \Aall \right]}{\phi_{k}^{\ms{(d)}}}.
\end{align*}

\noindent \textbf{E-step.}
The updates require computing $\varphi^{\ms{(d)}}_{\mathbf{i}k} := \Exp[\Ad_{\mathbf{i}k} \mid \Aall]$ and $\varphi_{\mathbf{i}ick}^{\ms{(d)}} := \Exp[\Ad_{\mathbf{i}ick}\mid \Aall]$. By multinomial thinning,  
\[
\Ad_{i_1 \dots i_d k} \mid \Ad_{i_1 \dots i_d} = a \sim \text{Multinomial}\left(a, \frac{\gamma_{k}^{\ms{(d)}}\sprod_{r=1}^d m_{i_r k}}{\sum_{k'=1}^K \gamma_{k'}^{\ms{(d)}}\sprod_{r=1}^d m_{i_r k'}}\right).
\]
Therefore, 
\[
\Exp[\Ad_{\mathbf{i}k} \mid \Aall]  = \Exp[\Ad_{\mathbf{i}k} \mid \Ad_{\mathbf{i}}] = \Ad_{\mathbf{i}} \cdot \frac{\gamma_{k}^{\ms{(d)}}\sprod_{r=1}^d m_{i_r k}}{\sum_{k'=1}^K \gamma_{k'}^{\ms{(d)}}\sprod_{r=1}^d m_{i_r k'}}.\]
To compute $\Exp[\Ad_{\mathbf{i}ick} \mid \Aall]$, we use the law of total expectation.
\begin{align*}
    \Exp[\Ad_{\mathbf{i}ick} \mid \Aall] &= \Exp[ \Exp[\Ad_{\mathbf{i}ick} \mid \Ad_{\mathbf{i}k},  \Ad_{\mathbf{i}}] \mid \Ad_{\mathbf{i}}]\\
    &= \Exp[ \Exp[\Ad_{\mathbf{i}ick} \mid \Ad_{\mathbf{i}k}] \mid \Ad_{\mathbf{i}}].
\end{align*}
$(\Ad_{\mathbf{i}ick})_{c=1}^C \mid \Ad_{\mathbf{i}k} \sim \text{Multinomial}\left(\Ad_{\mathbf{i}k}, \frac{\theta_{ic}\mathrm{w}_{ck}}{\sum_{c'=1}^C \theta_{ic'} \mathrm{w}_{kc'}}\right)$. As such, 
\begin{align*}
    \Exp[\Ad_{\mathbf{i}ick} \mid \Aall] &= \Exp[\Ad_{\mathbf{i}k} \mid \Ad_{\mathbf{i}}] \cdot \frac{\theta_{ic}\mathrm{w}_{ck}}{\sum_{c'=1}^C \theta_{ic'} \mathrm{w}_{c'k}}.
\end{align*}
We compute expectations in a compositional manner, first computing $\Exp [\Ad_{\mathbf{i} k} \mid \Ad_{\mathbf{i}}]$, and then $\Exp[\Ad_{\mathbf{i}ick} \mid \Ad_{\mathbf{i}}]$ conditional on $\Exp [\Ad_{\mathbf{i} k} \mid \Ad_{\mathbf{i}}]$.

\end{widetext}
\end{document}